\documentclass[conference,a4paper]{IEEEtran}
\usepackage[dvips]{graphicx,color}
\usepackage{latexsym}
\usepackage{array}

\begin{document}

%
%
\newcommand{\bc}{\begin{center}}  %
\newcommand{\ec}{\end{center}}
\newcommand{\befi}{\begin{figure}[h]}  %
\newcommand{\enfi}{\end{figure}}
\newcommand{\bsb}{\begin{shadebox}\begin{center}}   %
\newcommand{\esb}{\end{center}\end{shadebox}}
\newcommand{\bs}{\begin{screen}}     %
\newcommand{\es}{\end{screen}}
\newcommand{\bib}{\begin{itembox}}   %
\newcommand{\eib}{\end{itembox}}
\newcommand{\bit}{\begin{itemize}}   %
\newcommand{\eit}{\end{itemize}}
\newcommand{\defeq}{\stackrel{\triangle}{=}}
\newcommand{\qed}{\hbox{\rule[-2pt]{3pt}{6pt}}}
\newcommand{\beq}{\begin{equation}}
\newcommand{\eeq}{\end{equation}}
\newcommand{\beqa}{\begin{eqnarray}}
\newcommand{\eeqa}{\end{eqnarray}}
\newcommand{\beqno}{\begin{eqnarray*}}
\newcommand{\eeqno}{\end{eqnarray*}}
\newcommand{\ba}{\begin{array}}
\newcommand{\ea}{\end{array}}
\newcommand{\vc}[1]{\mbox{\boldmath $#1$}}
\newcommand{\lvc}[1]{\mbox{\footnotesize\boldmath $#1$}}

\newcommand{\wh}{\widehat}
\newcommand{\wt}{\widetilde}
\newcommand{\ts}{\textstyle}
\newcommand{\ds}{\displaystyle}
\newcommand{\scs}{\scriptstyle}
\newcommand{\vep}{\varepsilon}
\newcommand{\rhp}{\rightharpoonup}
\newcommand{\cl}{\circ\!\!\!\!\!-}
\newcommand{\bcs}{\dot{\,}.\dot{\,}}
\newcommand{\eqv}{\Leftrightarrow}
\newcommand{\leqv}{\Longleftrightarrow}
\newcommand{\MEq}[1]{\stackrel{
{\rm (#1)}}{=}}
\newcommand{\MLeq}[1]{\stackrel{
{\rm (#1)}}{\leq}}
\newcommand{\ML}[1]{\stackrel{
{\rm (#1)}}{<}}
\newcommand{\MGeq}[1]{\stackrel{
{\rm (#1)}}{\geq}}
\newcommand{\MG}[1]{\stackrel{
{\rm (#1)}}{>}}
\newcommand{\MPreq}[1]{\stackrel{
{\rm (#1)}}{\preceq}}
\newcommand{\MSueq}[1]{\stackrel{
{\rm (#1)}}{\succeq}}
\newtheorem{co}{Corollary} 
\newtheorem{lm}{Lemma} 
\newtheorem{Ex}{Example} 
\newtheorem{Th}{Theorem}
\newtheorem{df}{Definition} 
\newtheorem{pr}{Property} 
\newtheorem{pro}{Proposition} 
\newtheorem{rem}{Remark} 
\newcommand{\lcv}{convex } 
\newcommand{\hugel}{{\arraycolsep 0mm
                    \left\{\ba{l}{\,}\\{\,}\ea\right.\!\!}}
\newcommand{\Hugel}{{\arraycolsep 0mm
                    \left\{\ba{l}{\,}\\{\,}\\{\,}\ea\right.\!\!}}
\newcommand{\HUgel}{{\arraycolsep 0mm
                    \left\{\ba{l}{\,}\\{\,}\\{\,}\vspace{-1mm}
                    \\{\,}\ea\right.\!\!}}
\newcommand{\huger}{{\arraycolsep 0mm
                    \left.\ba{l}{\,}\\{\,}\ea\!\!\right\}}}
\newcommand{\Huger}{{\arraycolsep 0mm
                    \left.\ba{l}{\,}\\{\,}\\{\,}\ea\!\!\right\}}}
\newcommand{\HUger}{{\arraycolsep 0mm
                    \left.\ba{l}{\,}\\{\,}\\{\,}\vspace{-1mm}
                    \\{\,}\ea\!\!\right\}}}

\newcommand{\hugebl}{{\arraycolsep 0mm
                    \left[\ba{l}{\,}\\{\,}\ea\right.\!\!}}
\newcommand{\Hugebl}{{\arraycolsep 0mm
                    \left[\ba{l}{\,}\\{\,}\\{\,}\ea\right.\!\!}}
\newcommand{\HUgebl}{{\arraycolsep 0mm
                    \left[\ba{l}{\,}\\{\,}\\{\,}\vspace{-1mm}
                    \\{\,}\ea\right.\!\!}}
\newcommand{\hugebr}{{\arraycolsep 0mm
                    \left.\ba{l}{\,}\\{\,}\ea\!\!\right]}}
\newcommand{\Hugebr}{{\arraycolsep 0mm
                    \left.\ba{l}{\,}\\{\,}\\{\,}\ea\!\!\right]}}
\newcommand{\HUgebr}{{\arraycolsep 0mm
                    \left.\ba{l}{\,}\\{\,}\\{\,}\vspace{-1mm}
                    \\{\,}\ea\!\!\right]}}

\newcommand{\hugecl}{{\arraycolsep 0mm
                    \left(\ba{l}{\,}\\{\,}\ea\right.\!\!}}
\newcommand{\Hugecl}{{\arraycolsep 0mm
                    \left(\ba{l}{\,}\\{\,}\\{\,}\ea\right.\!\!}}
\newcommand{\hugecr}{{\arraycolsep 0mm
                    \left.\ba{l}{\,}\\{\,}\ea\!\!\right)}}
\newcommand{\Hugecr}{{\arraycolsep 0mm
                    \left.\ba{l}{\,}\\{\,}\\{\,}\ea\!\!\right)}}
\newcommand{\hugepl}{{\arraycolsep 0mm
                    \left|\ba{l}{\,}\\{\,}\ea\right.\!\!}}
\newcommand{\Hugepl}{{\arraycolsep 0mm
                    \left|\ba{l}{\,}\\{\,}\\{\,}\ea\right.\!\!}}
\newcommand{\hugepr}{{\arraycolsep 0mm
                    \left.\ba{l}{\,}\\{\,}\ea\!\!\right|}}
\newcommand{\Hugepr}{{\arraycolsep 0mm
                    \left.\ba{l}{\,}\\{\,}\\{\,}\ea\!\!\right|}}
\newenvironment{jenumerate}
	{\begin{enumerate}\itemsep=-0.25em \parindent=1zw}{\end{enumerate}}
\newenvironment{jdescription}
	{\begin{description}\itemsep=-0.25em \parindent=1zw}{\end{description}}
\newenvironment{jitemize}
	{\begin{itemize}\itemsep=-0.25em \parindent=1zw}{\end{itemize}}
\renewcommand{\labelitemii}{$\cdot$}
%
%
\newcommand{\iro}[2]{{\color[named]{#1}#2\normalcolor}}
%
 \newcommand{\irr}[1]{{\color[named]{Black}#1\normalcolor}}
%
\newcommand{\irg}[1]{{\color[named]{Green}#1\normalcolor}}
\newcommand{\irb}[1]{{\color[named]{Blue}#1\normalcolor}}
\newcommand{\irBl}[1]{{\color[named]{Black}#1\normalcolor}}
\newcommand{\irWh}[1]{{\color[named]{White}#1\normalcolor}}
%
 \newcommand{\irPk}[1]{{\color[named]{Black}#1\normalcolor}}
%
\newcommand{\irCb}[1]{{\color[named]{CadetBlue}#1\normalcolor}}

\newcommand{\irdg}[1]{{\color[named]{DarkGreen}#1\normalcolor}}
\newcommand{\irp}[1]{{\color[named]{Yellow}#1\normalcolor}}
\newcommand{\irO}[1]{{\color[named]{Orange}#1\normalcolor}}
\newcommand{\irBr}[1]{{\color[named]{Purple}#1\normalcolor}}
\newcommand{\irBw}[1]{{\color[named]{Brown}#1\normalcolor}}
\newcommand{\IrBr}[1]{{\color[named]{Purple}#1\normalcolor}}

\newcommand{\irMho}[1]{{\color[named]{Mahogany}#1\normalcolor}}
 \newcommand{\irOlg}[1]{{\color[named]{Black}#1\normalcolor}}
\newcommand{\irBg}[1]{{\color[named]{BlueGreen}#1\normalcolor}}
\newcommand{\irCy}[1]{{\color[named]{Cyan}#1\normalcolor}}
\newcommand{\irRyp }[1]{{\color[named]{RoyalPurple}#1\normalcolor}}

\newcommand{\irAqm}[1]{{\color[named]{Aquamarine}#1\normalcolor}}
\newcommand{\irRyb}[1]{{\color[named]{RoyalBule}#1\normalcolor}}
\newcommand{\irNvb}[1]{{\color[named]{NavyBlue}#1\normalcolor}}
\newcommand{\irSkb}[1]{{\color[named]{SkyBlue}#1\normalcolor}}
\newcommand{\irTeb}[1]{{\color[named]{TeaBlue}#1\normalcolor}}
\newcommand{\irSep}[1]{{\color[named]{Sepia}#1\normalcolor}}
\newcommand{\irReo}[1]{{\color[named]{RedOrange}#1\normalcolor}}
\newcommand{\irRur}[1]{{\color[named]{RubineRed}#1\normalcolor}}
\newcommand{\irSa }[1]{{\color[named]{Salmon}#1\normalcolor}}
\newcommand{\irAp}[1]{{\color[named]{Apricot}#1\normalcolor}}
%
%
\newenvironment{indention}[1]{\par
\addtolength{\leftskip}{#1}\begingroup}{\endgroup\par}
%
\newcommand{\namelistlabel}[1]{\mbox{#1}\hfill} 
\newenvironment{namelist}[1]{%
\begin{list}{}
{\let\makelabel\namelistlabel
\settowidth{\labelwidth}{#1}
\setlength{\leftmargin}{1.1\labelwidth}}
}{%
\end{list}}
%
%
\newcommand{\bfig}{\begin{figure}[t]}
\newcommand{\efig}{\end{figure}}
\setcounter{page}{1}

\newtheorem{theorem}{Theorem}

\newcommand{\ep}{\mbox{\rm e}}

\newcommand{\Exp}{\exp
}
\newcommand{\idenc}{{\varphi}_n}
\newcommand{\resenc}{\tilde{\varphi}}
\newcommand{\ID}{\mbox{\scriptsize ID}}
\newcommand{\TR}{\mbox{\scriptsize TR}}

\newcommand{\Av}{\mbox{\textsf E}}

\newcommand{\Vl}{|}
\newcommand{\Avw}[1]{({#1}|W^n)}

\newcommand{\Ag}{(R_1,R_2,P_{X^n},P_{Y^n}\Vl W^n)}
\newcommand{\Agv}[2]{({#1},{#2},P_{X^n},P_{Y^n}\Vl W^n)}

\newcommand{\Aga}{(R_1,P_{X^n},P_{Y^n}\Vl W^n)}
\newcommand{\Agva}[1]{({#1},P_{X^n},P_{Y^n}\Vl W^n)}

\newcommand{\Agb}{(R_2,P_{X^n},P_{Y^n}\Vl W^n)}
\newcommand{\Agvb}[1]{({#1},P_{X^n},P_{Y^n}\Vl W^n)}

\newcommand{\Jd}{X^nY^n}
\newcommand{\IdR}{r_n}

\newcommand{\Index}{{n,i}}

\newcommand{\cid} { {\cal C}_{\mbox{\scriptsize ID}}}
\newcommand{\cida}{ {\cal C}_{\mbox{{\scriptsize ID,a}}}}

\newcommand{\Wid}
{\hspace*{-4mm}}




\sloppy
\title{
Outer Bound of the Capacity Region 
for Identification via Multiple Access Channels
} 

\author{
\IEEEauthorblockN{
Yasutada Oohama
}
  \IEEEauthorblockA{
    University of Electro-Communications \\
    Tokyo, Japan\\
    Email: oohama@uec.ac.jp} 
}



\maketitle

\begin{abstract}
In this paper we consider the identification (ID) via multiple access 
channels (MACs). In the general MAC the ID capacity region includes the 
ordinary transmission (TR) capacity region. In this paper we discuss 
the converse coding theorem. We estimate two types of error 
probabilities of identification for rates outside capacity region, 
deriving some function which serves as a lower bound of the sum of two 
error probabilities of identification. This function has a property that 
it tends to zero as $n\to \infty$ for noisy channels satisfying the 
strong converse property. Using this property, we establish that  the 
transmission capacity region is equal to the ID capacity for the MAC 
satisfying the strong converse property. To derive the result 
we introduce a new resolvability problem on the output from the MAC.
We further develop a new method of converting the direct coding theorem 
for the above MAC resolvability problem into the converse coding theorem 
for the ID via MACs. 
%
\end{abstract}

\section{Introduction}

\arraycolsep 0.5mm

In 1989, Ahlswede and Dueck \cite{ad},\cite{ad2}, proposed a new 
framework of communication system using noisy channels. Their proposed 
framework called the identification via channels (or briefly say the ID 
channel) has opened a new and fertile area in the Shannon theory. After 
their pioneering work, the ID channel coding problem has intensively 
been studied from both theoretical and practical point of view 
(\cite{hv1}-\cite{ohIdch}). Identification via multi-way channels is an 
interesting problem. This problem was studied by \cite{AZ}, \cite{stb}, 
\cite{VM} and \cite{AV}. In spite of its theoretical interest and 
practical importance, the number of works on this theme seems to be 
relatively few.

In this paper we deal with the identification via multiple access 
channels (MACs) for general noisy channels with two inputs and one 
output finite sets and channel transition probabilities that may be 
arbitrary for every block length $n$. Steinberg \cite{stb}, and the 
author studied  the identification(ID) capacity region for general MACs. 
However, these works have a common gap in the proofs of the converse 
coding theorems. This gap was pointed out by Hayashi \cite{hy} and is 
not resolved yet. 

According to Steinberg \cite{stb}, by a similar argument to 
the case of single user 
channels we can show that the ID capacity region contains the 
transmission(TR) capacity region for the general MAC. He studied the 
converse coding theorem by using a lemma used to prove the converse 
coding theorem for the ID via single user channels. In this paper we 
focus on our attention to the converse coding theorem and study it by an 
approach different from that of Steinberg. We estimate two 
types of error probabilities of identification for rates outside 
capacity region, deriving some function which serves as a lower bound of 
the sum of two error probabilities of identification. This function has 
a property that it tends to zero as $n\to \infty$ for noisy channels 
satisfying the strong converse property. Using this property, we 
establish that  the transmission capacity region is equal to the ID 
capacity for the MAC satisfying the strong converse property. 

To derive the converse coding theorem for the ID channel Han and
Verd\'u \cite{hv2} introduced an approximation problem of output
distributions from single user channels. They call this problem
channel resolvability problem. They first proved a direct coding
theorem for the channel coding theorem and next proved a converse
coding theorem for the ID channel by converting the direct coding
theorem for the channel resolvability problem into the converse coding
theorem for the ID channel. To prove the converse coding theorem for
the ID via MACs, we formulate a new approximation problem of output
distributions from MACs. This problem is regard as a MAC resolvability
problem. A similar resolvability problem using MACs was studied 
by Steinberg \cite{stb2}. Our problem is some variant of his problem. 
We first establish a stronger result on the direct coding 
theorem for this problem by deriving an upper bound for the
approximation error of channel outputs to tend to zero as $n$ goes to
infinity. Next, we prove the converse coding theorem by converting the
direct coding theorem for the MAC resolvability problem into the
converse coding theorem for the ID via MACs.
\newcommand{\Emptyaaaa}{
The 2013 IEEE International Symposium on Information Theory will be
held at ICEC Convention Center near Taksim Square at the heart of
Istanbul, Turkey, from Sunday, July 7 to Friday, July 12, 2013.

\section{Submission}
Paper submission is handled online using the EDAS system:
\begin{center}
  https://edas.info/newPaper.php?c=13285
\end{center}
The papers are restricted in length to \textbf{five pages} in the
IEEEtran-conference style as presented here (including figures,
references, etc.).

Each paper must be classified as ``Eligible for student paper award''
or ``Not eligible for student paper award''. Paper that are selected
to be eligible for the student paper award should also contain
\emph{``THIS PAPER IS ELIGIBLE FOR THE STUDENT PAPER AWARD''} as a
first line in the abstract of the submission. Note that this reference
must be removed again in the final manuscript!

The deadline for the submission is \textbf{January 27, 2013}.
}

\section{Identification via Multiple Access Channels}
\vspace{-1mm}

Let ${\cal X}$, ${\cal Y}$ and ${\cal Z}$ be finite sets. Let ${\cal
P}({\cal X}^n)$ and ${\cal P}({\cal Y}^n)$ be sets of probability
distributions on ${\cal X}^n$ and ${\cal Y}^n$, respectively.  A
source ${\vc X}$ with alphabet ${\cal X}$ is the sequence $\{P_X^n:
P_X^n\in {\cal P}({\cal X}^n)\}_{n=1}^{\infty}$ and a source ${\vc Y}$
with alphabet ${\cal Y}$ is the sequence $\{P_Y^n: P_Y^n\in {\cal
P}({\cal Y}^n)\}_{n=1}^{\infty}$. Similarly, a noisy channel {\vc W}
with two inputs alphabets ${\cal X}$ and ${\cal Y}$ and one output
alphabet ${\cal Z}$ is a sequence of conditional distributions
$\{W^n(\cdot|\cdot,\cdot)\}_{n=1}^{\infty}$, where
$W^n(\cdot|\cdot,\cdot)=\{$ $W^n(\cdot|{\vc x}, {\vc y})\in{\cal
P}({\cal Z}^n)$ $\}_{({\lvc x},{\lvc y})\in{\cal X}^n\times {\cal
Y}^n}$. Next, for $P_{X^n}\in {\cal P}({\cal X}^n)$, $P_{Y^n}\in {\cal
P}({\cal Y}^n)$ and ${\vc z}\in {\cal Z}^n$, set 
\vspace{-2mm}
\beqa 
&&P_{X^n}P_{Y^n}W^n({\vc z}) \nonumber\\ &=&\sum_{ ({\lvc x}, {\lvc
y})\in {\cal X}^n\times {\cal Y}^n } P_{X^n}({\vc x})P_{Y^n}({\vc y})
W^n({\vc z}|{\vc x},{\vc y})\,, 
\eeqa 
which becomes a probability distribution on ${\cal Z}^n$. 
We denote it by $P_{X^n}P_{Y^n}W^n=$
$\{P_{X^n}P_{Y^n}$ $W^n({\vc z})$ $\}_{{\lvc z}\in{\cal Z}^n}$. Set
$P_{Z^n}=P_{X^n}P_{Y^n}W^n$ and call $P_{Z^n}$ {\it the response of
$(P_{X^n},P_{Y^n})$ through noisy channel $W^n$ (or briefly the
response of $(P_{X^n},P_{Y^n})$)}.

An $(n,N_1,N_2,\mu_n,\lambda_n)$ ID code for $W^n$ is a collection
$\{(P_{X^n|i}, P_{Y^n|j}, $ $D_{i,j}),$ $i=1,2,\cdots, N_1,$
$j=1,2,\cdots,N_2\}$ such that
$$
\ba{ll}
\mbox{1)}& P_{X^n|i}\in {\cal P}({\cal X}^n)\,, 
           P_{Y^n|j}\in {\cal P}({\cal Y}^n)\,, 
\vspace{1mm}\\    
\mbox{2)}& { D}_{i,j} \subseteq {\cal Z}^n\,,
\vspace{1mm}\\    
\mbox{3)}& P_{Z^n|i,j} \mbox{ is the response of }
           (P_{X^n|i}, P_{Y^n|j})\,,
\vspace{1mm}\\    
\mbox{4)}& \ds \mu_{n,ij}=P_{Z^n|i,j} ({D}_{i,j}^c)\,,
           \ds \mu_n= \max_{\scs 1\leq i \leq N_1\,,
               \atop{\scs 1\leq j \leq N_2} }\mu_{n,ij}\,,
\vspace{1mm}\\
\mbox{5)}& \ds \lambda_{n,ij}=
           \ds\max_{(k,l) \ne (i,j)}
               P_{Z^n|k,l}({ D}_{i,j} )\,,
           \ds\lambda_n=\max_{\scs 1\leq i \leq N_1\,,
           \atop{ \scs 1\leq j \leq N_2}}
           \lambda_{n,ij}\,. \\
\ea
$$
The rate of an $(n,N_1,N_2,\mu_n,\lambda_n)$ ID code is defined by 
$$
r_{i,n}\defeq\frac{1}{n}\log\log N_i\,,i=1,2. 
$$

A rate pair $(R_1,R_2)$ is said to be $(\mu,\lambda)$-achievable ID
rate pair if there exists an $(n,N_1,$ $N_2,$ $\mu_n,$ $\lambda_n)$
code such that
$$
\left.
\ba{l}
\ds\limsup_{n\to\infty}\mu_n\leq \mu\,, 
\ds\limsup_{n\to\infty}\lambda_n\leq \lambda\,,
\vspace{1mm}\\
\ds\liminf_{n\to\infty} r_{i,n}\geq R_i\,,i=1,2\, .
\ea
\right\}
$$
The set of all $(\mu,\lambda)$-achievable ID rate pairs for ${\vc W}$
is denoted by ${\cal C}_{\mbox{\scriptsize ID}}(\mu,\lambda|{\vc W})$, 
which we call the $(\mu,\lambda)$-ID capacity region.   

To state results for the identification capacity region, we prepare 
several quantities which are defined based on the notion of 
the {\it information} {\it spectrum} introduced 
by Han and Verd\'u \cite{hv2}.   

\begin{df}{\rm For $n=1,2,\cdots$, let $X^n$ and $Y^n$ be an arbitrary 
prescribed independent random variable 
taking values in ${\cal X}^n$ 
and ${\cal Y}^n$, respectively. 
The probability mass function of $X^n$ and $Y^n$ is 
$P_{X^n}({\vc x}),$ ${\vc x}\in {\cal X}^n$ and 
$P_{Y^n}({\vc x}),$ ${\vc y}\in {\cal Y}^n$, respectively.
A pair of two independent sources $({\vc X},{\vc Y})$ with alphabet 
${\cal X}\times {\cal Y}$ is the sequence $\{(P_{X^n},$ $P_{Y^n}):$ 
$P_{X^n}\in {\cal P}({\cal X}^n),$ $P_{Y^n}\in {\cal P}({\cal Y}^n)\}$. 
A collection of such $({\vc X},{\vc Y})$ is denoted by ${\cal S}_I$.   
Let $Z^n$ be an output random variable when we use 
$X^n$ and ${Y}^n$ as two inputs of the noisy channel $W^n$. 
In this case the joint probability mass function
of $(X^n,Y^n, Z^n)$ 
denoted by $P_{X^nY^nZ^n}({\vc x}, {\vc y},$ 
${\vc z}),$ 
$({\vc x}, {\vc y}, {\vc z})$ $\in {\cal X}^n\times {\cal Y}^n 
\times {\cal Z}^n$ is equal to 
$P_{X^n}({\vc x})$$P_{Y^n}({\vc y})$
$W^n({\vc z}|{\vc x},{\vc y})$.

}
\end{df}

\begin{df}{\rm  Given a joint distribution 
$P_{X^nY^nZ^n}($ ${\vc x},$ ${\vc y},$ ${\vc z})=$ 
$P_{X^n}({\vc x})$$P_{Y^n}({\vc y})$ $W^n({\vc
z}|{\vc x},{\vc y})$, the information density 
is the function defined on ${\cal X}^n\times {\cal Y}^n:$ 
\beqno 
i_{X^nY^nZ^n}({\vc x};{\vc z}|{\vc y})
&=&\log\frac{W^n({\vc z}|{\vc x},{\vc y})}
 {P_{Z^n|Y^n}({\vc z}|{\vc y})}\,,\\
i_{X^nY^nZ^n}({\vc y};{\vc z}|{\vc x})
&=&\log\frac{W^n({\vc z}|{\vc x},{\vc y})}
 {P_{Z^n|X^n}({\vc z}|{\vc x})}\,,\\
i_{X^nY^nZ^n}({\vc x}{\vc y}; {\vc z})
&=&\log\frac{W^n({\vc z}|{\vc x},{\vc y})}
     {P_{Z^n}({\vc z})}\,.  
\eeqno
}\end{df}

\begin{df}
{\rm
Let $\{A_n\}_{n=1}^{\infty}$ be a sequence of arbitrary 
real-valued random variables. We introduce 
the notion of the so-called {\it probabilistic} {\it limsup/inf} 
in the following.  
\beqno
\vspace*{-3mm}\mbox{\rm p-}\limsup_{n\to\infty}A_n 
&\defeq & 
\inf\{
\alpha : \lim_{n\to\infty}\Pr\{A_n \geq\alpha\}=0
\}\,,
\\
\vspace*{-3mm}\mbox{\rm p-}\liminf_{n\to\infty}A_n  
&\defeq& 
\sup\{
\alpha : \lim_{n\to\infty}\Pr\{A_n\leq \alpha\}=0
\}\,.
\eeqno
}
\end{df}

The {\it probabilistic limsup/inf} in the above definitions is
considered as an extension of ordinary (deterministic) liminf. The
operation of limsup/inf has the same properties as those of the
operation of limsup/inf. For the details see Han and Verd\'u \cite{hv2} 
and Han \cite{han}. 

\begin{df}{\rm
Set 
\beqno
\underline{I}({\vc X};{\vc Z}|{\vc Y})
&\defeq&\mbox{p-}\liminf\frac{1}{n} i_{X^nY^nZ^n}(X^n;Z^n|Y^n),
\\
\underline{I}({\vc Y}; {\vc Z}|{\vc X})
&\defeq&\mbox{p-}\liminf\frac{1}{n} i_{X^nY^nZ^n}(Y^n;Z^n|X^n),
\\
\underline{I}({\vc X}{\vc Y};{\vc Z})
&\defeq&\mbox{p-}\liminf\frac{1}{n} i_{X^nY^nZ^n}(X^nY^n;Z^n).
\eeqno
Furthermore, set  
\beqno
\underline{\cal C}({\vc X},{\vc Y} \Vl {\vc W})
\defeq\left\{(R_1,R_2): \right.
&&\hspace*{-2mm}R_1\leq\underline{I}({\vc X};{\vc Z}|{\vc Y}),
\nonumber\\   
&&\hspace*{-2mm}R_2\leq\underline{I}({\vc Y}; {\vc Z}|{\vc X}),
\nonumber\\   
&&\hspace*{-2mm}R_1+R_2\leq\underline{I}({\vc X}{\vc Y};{\vc Z})
\left.\right\},
\eeqno
$$
\underline{\cal C}({\vc W})
\defeq \bigcup_{(\lvc X,\lvc Y )\in {\cal S}_I}
\underline{\cal C}({\vc X},{\vc Y} \Vl {\vc W}).
$$
}
{\rm Set 
\beqno
\overline{I}({\vc X};{\vc Z}|{\vc Y})
&\defeq&\mbox{p-}\limsup\frac{1}{n} i_{X^nY^nZ^n}(X^n;Z^n|Y^n),
\\
\overline{I}({\vc Y}; {\vc Z}|{\vc X})
&\defeq&\mbox{p-}\limsup\frac{1}{n} i_{X^nY^nZ^n}(Y^n;Z^n|X^n),
\\
\overline{I}({\vc X}{\vc Y};{\vc Z})
&\defeq&\mbox{p-}\limsup\frac{1}{n} i_{X^nY^nZ^n}(X^nY^n;Z^n).
\eeqno
Furthermore, set  
\beqno
\overline{\cal C}({\vc X},{\vc Y} \Vl {\vc W})
\defeq\left\{(R_1,R_2): \right.
&&\hspace*{-2mm}R_1\leq\overline{I}({\vc X};{\vc Z}|{\vc Y}),
\nonumber\\   
&&\hspace*{-2mm}R_2\leq\overline{I}({\vc Y}; {\vc Z}|{\vc X}),
\nonumber\\   
&&\hspace*{-2mm}R_1+R_2\leq\overline{I}({\vc X}{\vc Y};{\vc Z})
\left.\right\},   
\eeqno
$$
\overline{\cal C}({\vc W})
\defeq\bigcup_{({\lvc X},{\lvc Y} )\in {\cal S}_I}
\overline{\cal C}({\vc X},{\vc Y} \Vl {\vc W}).
$$
}
\end{df}

Han \cite{han},\cite{hanMac} proved that $\underline{\cal C}({\vc W})$ 
is equal to the ordinary transmission capacity region 
for general MACs. Han \cite{hanMac} proved that
when $\underline{\cal C}({\vc W})$ 
$=\overline{\cal C}({\vc W})$, the strong converse 
property holds, i.e., the error probability 
of transmission goes to one as $n\to \infty$ for 
all transmission rates outside the capacity region.    


Identification via multiple access channels was first 
investigated by Steinberg \cite{stb}. His result 
is the following. 
%

{\it Theorem A (Steinberg \cite{stb}) } \label{th:th1a}
{\rm For general noisy channel ${\vc W}$, we have  
\beq
{\cal C}_{\ID}(0,0|{\vc W})\supseteq\underline{\cal C}({\vc W})\,.
\eeq
\label{th:th0-1}
}

The above theorem can be proved by an argument quite similar to the case 
of the identification via single-user channels. Steinberg \cite{stb} also 
studied the converse coding theorem. In \cite{stb} he established a new 
lemma useful to prove the converse coding theorem of the identification 
via single-user channels. Using this lemma and the capacity formula 
by Verd\'u \cite{Ved}, he obtained a result on the converse coding theorem 
for the identification via MACs.

%
%
%
%


In this paper we study the converse coding theorem for the 
ID via general MACs. Our approach is different from 
that of Steinberg \cite{stb}. 
We derive a function which 
serves as an upper bound of $1-\mu_n-\lambda_n$ for 
general MACs. To obtain this result we formulate a new resolvability
problem for the general MAC, that is, an approximation problem of
output random variables via MACs. We consider this problem and derive
an upper bound of the approximation error. This upper bound is useful
for analyzing the error probability of identification outside the ID
capacity region.


\section{Main Results}

\subsection{Definitions of Functions and their Properties} 
We first define several functions to describe our results 
and state their basic properties. 
\begin{df}{\rm Let $S$ be an arbitrary subset of 
${\cal X}^n\times$ 
${\cal Y}^n\times$ 
${\cal Z}^n$ 
and ${\vc 1}_{S}({\vc x},{\vc y},{\vc z})$
be indicator functions which takes value one on $S$ and zero
outside $S$. 
Set
\beqno
\zeta_{n,1,S}&=&\zeta_{n,1,S}(R_1,P_{X^n},P_{Y^n} \Vl W^n)
\nonumber\\
&=&\Av
\left[{\ep}^{-n[R_1-\frac{1}{n}i_{X^nY^nZ^n}(X^n;Z^n|Y^n)]}
\right.
\nonumber\\
&&\qquad\qquad\qquad\qquad\qquad\times
\Bigl.{\vc 1}_{S}(X^n,Y^n,Z^n)\Bigr],
\nonumber\\
\zeta_{n,2,S}&=& \zeta_{n,2,S}(R_2,P_{X^n},P_{Y^n} \Vl W^n)
\nonumber\\
&=&\Av\left[
         {\ep}^{-n[R_2-\frac{1}{n}i_{X^nY^nZ^n}(Y^n;Z^n|X^n)]}
         \right.
\nonumber\\
&&       \qquad\qquad\qquad\qquad\qquad\times
         \Bigl.{\vc 1}_{S}(X^n,Y^n,Z^n)\Bigr],
\nonumber\\
\zeta_{n,3,S}&=&\zeta_{n,3,S}(R_1,R_2,P_{X^n},P_{Y^n},W^n)
\nonumber\\
&=&\Av\left[\left\{
          {\ep}^{-n[R_1-\frac{1}{n}i_{X^nZ^n}(X^n;Z^n)]}
          \right.\right.
\nonumber\\
&&\qquad+{\ep}^{-n[R_2-\frac{1}{n}i_{Y^nZ^n}(Y^n;Z^n)]}
\nonumber\\
&&\qquad+{\ep}^{-n[R_1+R_2-\frac{1}{n}i_{X^nY^nZ^n}(X^nY^n;Z^n)]}
         \Bigl.\Bigr\}
\nonumber\\
&&\qquad\qquad\qquad\qquad \Bigl.\times
          {\vc 1}_{S}(X^n,Y^n,Z^n)\Bigr]\,.
\eeqno
}\end{df}

\begin{df}{\rm
Set
\beqno
T_{\gamma}
&=&
\left\{\right.
({\vc x},{\vc y}, {\vc z})\in {\cal X}^n\times 
{\cal Y}^n\times {\cal Z}^n:
\nonumber\\
& &\quad\:\: \frac{1}{n}i_{X^nY^nZ^n}({\vc x};{\vc z}|{\vc y})
        \leq R_1-\gamma\,,%
\nonumber\\
&&\mbox{ or }\frac{1}{n}i_{X^nY^nZ^n}({\vc y};{\vc z}|{\vc x})
        \leq R_2-\gamma\,,%
\nonumber\\
& &\mbox{ or }\frac{1}{n}i_{X^nY^nZ^n}({\vc x}{\vc y};{\vc z})
        \leq  R_1+R_2-2\gamma%
\left.\right\}\,. 
\eeqno
Define three subsets of 
${\cal X}^n\times {\cal Y}^n\times {\cal Z}^n$ by 
\beqno
T_{1,\gamma}
&=&
\left\{\right.
({\vc x},{\vc y}, {\vc z})\in {\cal X}^n\times {\cal Y}^n\times {\cal Z}^n:
\nonumber\\
&&\quad \frac{1}{n}i_{X^nY^nZ^n}({\vc x};{\vc z}|{\vc y})\leq R_1-\gamma
\left.\right\}\,, 
\label{eqn:defT1}
\\
T_{2,\gamma}
&=&
\left\{\right.
({\vc x},{\vc y}, {\vc z})\in {\cal X}^n\times {\cal Y}^n\times {\cal Z}^n:
\nonumber\\
&&\quad\frac{1}{n}i_{X^nY^nZ^n}({\vc y};{\vc z}|{\vc x})\leq R_2-\gamma
\left.\right\}\,, 
\label{eqn:defT2}
\\
T_{3,\gamma}
&=&
\left\{\right.
({\vc x},{\vc y}, {\vc z})\in {\cal X}^n\times {\cal Y}^n\times {\cal Z}^n:
\nonumber\\
&&\quad
\frac{1}{n} i_{X^nZ^n}({\vc x};{\vc z})\leq R_1-\gamma\,,
\nonumber\\
&&\quad\frac{1}{n}i_{Y^nZ^n}({\vc y};{\vc z})\leq R_2-\gamma\,,
\nonumber\\
&&\quad\frac{1}{n}i_{X^nY^nZ^n}({\vc x}{\vc y};{\vc z})
\leq R_1+R_2 -2\gamma \left.\right\}\,. 
\label{eqn:defT3}
\eeqno
Set
\beqno
\lefteqn{\Omega_{n,i,\gamma}^{(1)}
(R_i,P_{X^n},P_{Y^n}\Vl W^n)}
\\
&=&\Pr\left\{(X^n,Y^n,Z^n)\notin T_{i,\gamma}\right\}, i=1,2,
\\
\lefteqn{\Omega_{n,3,\gamma}^{(1)}
(R_1,R_2,P_{X^n},P_{Y^n}\Vl W^n)}
\\
&=&\Pr\left\{(X^n,Y^n,Z^n)\notin T_{3,\gamma}\right\},
\\
\lefteqn{
\Omega_{n,i,\gamma}^{(2)}
(R_i,P_{X^n},P_{Y^n}\Vl W^n)}
\\
&=&\zeta_{n,i,T_{i,\gamma}}(R_i,P_{X^n},P_{Y^n}\Vl W^n), i=1,2,
\\
\lefteqn{
\Omega_{n,3,\gamma}^{(2)}
(R_1,R_2,P_{X^n},P_{Y^n}\Vl W^n)}
\\
&=&\zeta_{n,3,T_{3,\gamma}}(R_1,R_2,P_{X^n},P_{Y^n}\Vl W^n),
\\
\lefteqn{
\Omega_{n,i,\gamma}(R_i,P_{X^n},P_{Y^n}\Vl W^n)}
\nonumber\\
&=&4\Omega_{n,i,\gamma}^{(1)}(R_i,P_{X^n},P_{Y^n}\Vl W^n) 
\nonumber\\
& &\quad+3\sqrt{\Omega_{n,i,\gamma}^{(2)}(R_i,P_{X^n},P_{Y^n} \Vl W^n)}, i=1,2,
\nonumber\\
\lefteqn{
\Omega_{n,3,\gamma}\Ag}
\nonumber\\
&=&4\Omega_{n,3,\gamma}^{(1)}\Ag
\nonumber\\
& &\quad+3\sqrt{\Omega_{n,3,\gamma}^{(2)}\Ag}.
\eeqno
Furthermore, set
\beqno
\lefteqn{\hspace*{-10mm}
\Omega_{n,\gamma}(R_1,R_2, P_{X^n},P_{Y^n}\Vl W^n)}
\\
&=&\min\{
\Omega_{n,1,\gamma}(R_1,P_{X^n},P_{Y^n}\Vl W^n),
\\
& & \qquad \Omega_{n,2,\gamma}(R_2,P_{X^n},P_{Y^n}\Vl W^n),
\\
& &\qquad \Omega_{n,3,\gamma}(R_1,R_2,P_{X^n},P_{Y^n}\Vl W^n)\}
\eeqno
Finally, set
\beqa
& &\Omega_{n,\gamma}(R_1,R_2 \Vl W^n)
\nonumber\\
&=&\sup_{\scs (P_{X^n},P_{Y^n})
         \atop{\scs \in {\cal P}({\cal X}^n)\times {\cal P}({\cal Y}^n)}  
         }
\Omega_{n,\gamma}(R_1,R_2,P_{X^n},P_{Y^n} \Vl W^n)\,.
\eeqa
}\end{df}

We can prove that $\Omega_{n,\gamma}($ $R_1,R_2,$ $W^n)$ and 
$\Omega_{n,\gamma}($ $R_1,R_2$ $P_{X^n},$ $P_{Y^n},$ $W^n)$ 
satisfy the following two properties.
\begin{pr}\label{pr:pr0}{$\quad$
\begin{itemize}
\item[{\rm a)}] For any $0\leq \gamma <\tau$, 
\beqno
& &\Omega_{n,i,0}^{(1)}(R_i,P_{X^n},P_{Y^n}\Vl W^n)
\\
&=& \Omega_{n,i,\gamma}^{(1)}(R_i-\gamma,P_{X^n},P_{Y^n}\Vl W^n), i=1,2,
\\
& &\Omega_{n,3,0}^{(1)}(R_1,R_2, P_{X^n},P_{Y^n}\Vl W^n)
\\
&=& \Omega_{n,3,\gamma}^{(1)}
(R_1-\gamma,R_2-\gamma,P_{X^n},P_{Y^n}\Vl W^n),
\\
& &\Omega_{n,\gamma}^{(2)}(R_i,P_{X^n},P_{Y^n}\Vl W^n)
\\
&=& {\ep}^{-n\gamma} \Omega_{n,i,0}^{(2)}
(R_i-\gamma,P_{X^n},P_{Y^n}\Vl W^n), i=1,2,
\\
& &\Omega_{n,3,\gamma}^{(2)}(R_1,R_2, P_{X^n},P_{Y^n}\Vl W^n)
\\
&\leq & {\ep}^{-n\gamma} 
\Omega_{n,3,0}^{(2)}(R_1-\gamma,R_2-\gamma,P_{X^n},P_{Y^n}\Vl W^n),
\\
& &\Omega_{n,i,\gamma}^{(2)}(R_i,P_{X^n},P_{Y^n}\Vl W^n)   
\leq {\ep}^{-n\gamma}, i=1,2,
\\
& &\Omega_{n,3,\gamma}^{(2)}(R_1,R_2, P_{X^n},P_{Y^n}\Vl W^n)   
\leq 3{\ep}^{-n\gamma},
\\
& &\Omega_{n,i,\gamma}^{(2)}(R_i,P_{X^n},P_{Y^n}\Vl W^n)   
\\
&\leq&{\ep}^{-n\tau}
+\Omega_{n,i,\tau}^{(1)}(R_i,P_{X^n},P_{Y^n}\Vl W^n)   
\\
&  &-\Omega_{n,i,\gamma}^{(1)}(R_i, P_{X^n},P_{Y^n}\Vl W^n), i=1,2,   
\\
& &\Omega_{n,3,\gamma}^{(2)}(R_1,R_2, P_{X^n},P_{Y^n}\Vl W^n) 
\\
&\leq&3{\ep}^{-n\tau}
+\Omega_{n,3,\tau}^{(1)}(R_1,R_2, P_{X^n},P_{Y^n}\Vl W^n)   
\\
&  &-\Omega_{n,3,\gamma}^{(1)}(R_1,R_2, P_{X^n},P_{Y^n}\Vl W^n).   
\eeqno

\item[{\rm b)}] For any $\gamma\geq 0$ and $R_1\geq 0$, 
$R_2\geq 0$,
\beqno
&&0\leq\Omega_{n,i,\gamma}^{(1)}(R_i,P_{X^n},P_{Y^n}\Vl W^n)\leq 1, i=1,2,
\\
&&0\leq\Omega_{n,3,\gamma}^{(1)}(R_1,R_2,P_{X^n},P_{Y^n}\Vl W^n)\leq 1. 
\eeqno
\end{itemize}
}
\end{pr}

\begin{pr}\label{pr:pr1}{$\quad$
\rm
\begin{itemize}
\item[{\rm a)}] For any $\gamma\geq 0$ and $R_1, R_2\geq 0$,
      $$
      0\leq\Omega_{n,\gamma}(R_1,R_2, W^n)
      \leq \frac{73}{16}\,.
      $$ 
\item[{\rm b)}] Set
\beqno
& &\overline{\cal C}^{\prime}({\vc X}, {\vc Y}\Vl {\vc W})
\\
&\defeq&\overline{\cal C}({\vc X}, {\vc Y}\Vl {\vc W})
\\
& &\cup \{(R_1,R_2): 
R_1 \leq \overline{I}({\vc X};{\vc Z}),
R_2 \leq \overline{I}({\vc Y};{\vc Z}|{\vc X})\}
\\
& &
\cup \{(R_1,R_2): 
R_1 \leq \overline{I}({\vc X};{\vc Z}|{\vc Y}),
R_2 \leq \overline{I}({\vc Y};{\vc Z})\}
\\
&&\overline{\cal C}^{\prime}({\vc W})
\defeq\bigcup_{({\lvc X},{\lvc Y} )\in {\cal S}_I}
\overline{\cal C}^{\prime}({\vc X},{\vc Y} \Vl {\vc W}).
\eeqno 
It is obvious that 
$\overline{\cal C}({\vc W})\subseteq \overline{\cal C}^{\prime}({\vc W})$.
If $(R_1,R_2) \notin \overline{\cal C}^{\prime}({\vc W})$, then, 
      there exists a small positive number $\gamma_0$ such that for 
      any $\gamma\in [0,\gamma_0)$, 
      $$
      \lim_{n\to\infty}\Omega_{n,\gamma}(R_1,R_2\Vl W^n)=0.
      $$
\end{itemize}
}
\end{pr}

Proofs of Properties \ref{pr:pr0} and \ref{pr:pr1} are 
quite parallel with those of Properties 1 and 2 
in \cite{ohIdch}. Proof of Property 2 part b) 
is given in the appendix. 

\newcommand{\Emptyaac}{
Property \ref{pr:pr1} parts a) and b) are obvious from 
the definitions of 
$\Omega_{n,0}(R|W^n)$, 
$\Omega_{n,1,0}(R|W^n)$,
$\sigma_{n}(R|W^n)$, and $\sigma_{n,1}(R|W^n)$.  
Proof of Property \ref{pr:pr1} part c) will be given in Appendix A. 
Property \ref{pr:pr3} immediately follows from 
Property \ref{pr:pr1}. 
}

\subsection{Statement of Results} 

Our main result for the identification via MACs is the following.
\begin{pro}\label{pro:thM2} {\rm 
For any $(n, N_1,$ $N_2,$ $\mu_n,$ $\lambda_n)$ code with 
$\mu_n+$ $\lambda_n$  $ < 1$, if 
the rate $r_{i,n}=(1/n)\log\log N_i$ satisfies   
\beqa
r_{1,n} &\geq&  R_1+\frac{\log n}{n}
               +\frac{1}{n}\log\log(3|{\cal X}|)^2\,,
\label{eqn:rcnd}  
\\
r_{2,n} &\geq & R_2 +\frac{\log n}{n}
               +\frac{1}{n}\log\log(3|{\cal Y}|)^2\,,
\label{eqn:rcnd2}  
\eeqa
then, for any $\gamma\geq 0$, the sum $\mu_n+\lambda_n$ of two error 
probabilities satisfies the following: 
\beq
1-\mu_n-\lambda_n \leq \Omega_{n,\gamma}(R_1,R_2\Vl W^n)\,.
\eeq
}
\end{pro}

From this proposition, we obtain the following corollary.   
\begin{co} \label{co:co0}{\rm 
For any sequence of ID codes 
$\{(n,$ $N_1,$, $N_2,$ 
$\mu_n,$ $\lambda_n)$
$\}_{n=1}^{\infty}$ satisfying $\mu_n+$ $\lambda_n$ $ < 1$, 
$n=1,2,\cdots$, if 
$$
\liminf_{n\to\infty} r_{i,n} \geq R_i,\: i=1,2, 
$$
then, for any $\delta>0$, there exists $n_0=n_0(\delta)$ 
such that for $n\geq n_0$, 
\beq
1-{\mu}_n-{\lambda}_n \leq \Omega_{n,\gamma}(R_1-\delta, R_2-\delta\Vl W^n)\,.
\label{eqn:ieqco00}
\eeq
}
\end{co} 

It immediately follows from Theorem A, Corollary \ref{co:co0} and 
Property \ref{pr:pr1} part b) that the following strong 
converse theorem holds.  
%
%

\begin{Th}
\label{co:coHV}
For any sequence of ID codes $\{(n, N_1,$ $N_2,$ $\mu_n,$ $\lambda_n)$
$\}_{n=1}^{\infty}$ satisfying $\mu_n$ $+$ $\lambda_n$ $<1$, 
$n=1,2,\cdots$, if  
\beqno
\liminf_{n\to\infty} r_{i,n} \geq R_i, i=1,2, \quad  
(R_1,R_2) \notin \overline{\cal C}^{\prime}({\vc W}),
\eeqno
then, 
$$
\liminf_{n\to\infty}\{\mu_n+\lambda_n\}=1, 
$$
which implies that for any $\mu\geq 0, \lambda\geq 0,$
$\mu + \lambda < 1$ and any noisy channel ${\vc W}$, 
$$
\underline{\cal C}({\vc W})\subseteq 
{\cal C}_{\rm ID}(\mu,\lambda|{\vc W})\subseteq 
\overline{\cal C}^{\prime}({\vc W}).
$$
In particular, if 
$$ 
\underline{\cal C}({\vc W})
=\overline{\cal C}({\vc W})=\overline{\cal C}^{\prime}({\vc W}),
$$ 
then, for any $\mu\geq 0,\lambda \geq 0, $ $\mu+\lambda < 1$, 
$$
\underline{\cal C}({\vc W})={\cal C}_{\rm ID}(\mu,\lambda|{\vc W})
=\overline{\cal C}({\vc W})=\overline{\cal C}^{\prime}({\vc W}).
$$
Furthermore, $\mu_n+\lambda_n$ converges to one as $n\to\infty$ 
at rates above the ID capacity. This implies that 
the strong converse property holds with respect to the sum of 
two types of error probabilities.
\end{Th}

\subsection{Results for the Average Error Criterion}  

We have so far dealt with the case that the error probabilities of   
identification are measured in the maximum sense. In this subsection
we consider the following average error criterion:
\beq
\left.
\ba{rcl}
\bar{\mu}_n&=&\ds\frac{1}{N_1N_2}\sum_{i=1}^{N_1}
                              \sum_{j=1}^{N_2}
                              \mu_{n,ij}\,,\\
\bar{\lambda}_n&=&\ds\frac{1}{N_1N_2}\sum_{i=1}^{N_1}
                                  \sum_{j=1}^{N_2}
                                  \lambda_{n,ij}\,.
\ea
\right\}
\eeq
For $0\leq \mu, \lambda\leq 1$, let 
$\cida(\mu,\lambda|{\vc W})$ be denoted
by the identification capacity defined by replacing the maximum
error probability criterion by the above average error 
probability criterion.  
Since $\bar{\mu}_n\leq {\mu}_n$ and 
      $\bar{\lambda}_n\leq{\lambda}_n$, it is obvious that
for any $\mu,\lambda \geq 0$,  
\beq   
{\cal C}_{\ID}(\mu,\lambda|{\vc W})
\subseteq \cida(\mu,\lambda|{\vc W})\,.
\eeq

We shall show that $\cida(\mu,\lambda|{\vc W})$ has 
the same outer bound as $\cid (\mu,\lambda|{\vc W})$. 
An important key result in the case of the average 
error criterion is given in the following proposition. 

\begin{pro}\label{pro:thM3} {\rm Fix $\tau>0$ arbitrarily.  
For any $(n, N_1,$ $N_2,$ $\bar{\mu}_n,$ $\bar{\lambda}_n)$ code 
with $\bar{\mu}_n+$ $\bar{\lambda}_n$  $ < 1$
if the rate 
$r_{i,n}=\frac{1}{n}\log\log N_i$, $i=1,2$ satisfy   
\beqa
r_{1,n} 
&\geq & R_1 +\tau + \frac{\log n}{n}
+\frac{1}{n}\log\log |{\cal X}|^2\,,
\label{eqn:avrcnd1}  
\\
r_{2,n} 
&\geq & R_2 +\tau + \frac{\log n}{n}
+\frac{1}{n}\log\log |{\cal Y}|^2\,,
\label{eqn:avrcnd2}  
\eeqa
then, for any $\gamma\geq 0$, the sum 
$\bar{\mu}_n+\bar{\lambda}_n$ of two average 
error probabilities satisfies the following: 
\beqno
1-\bar{\mu}_n-\bar{\lambda}_n
&\leq& \Omega_{n,\gamma}(R_1,R_2\Vl W^n)
\nonumber\\
&   & +\nu_{n,\tau}(R_1,R_2,|{\cal X}|,|{\cal Y}|),
\eeqno
where
\beqno
&&\nu_{n,\tau}(R_1,R_2,|{\cal X}|, |{\cal Y}|)
\nonumber\\
&\defeq& 
    |{\cal X}|^{-2n({\ep}^{n\tau}-1){\ep}^{nR_1}}
    |{\cal Y}|^{ -2n({\ep}^{n\tau}-1){\ep}^{nR_2}}
\nonumber\\
& &+|{\cal X}|^{-2n({\ep}^{n\tau}-1){\ep}^{nR_1}}\cdot
             |{\cal Y}|^{ -2n({\ep}^{n\tau}-1){\ep}^{nR_2}}.
\eeqno
}
Since ${\ep}^{n\tau}-1\geq n\tau$, we have 
\beqno
0&\leq&\nu_{n,\tau}(R_1,R_2,|{\cal X}|, |{\cal Y}|)
\\
 &\leq& |{\cal X}|^{-{2n^2\tau}{\ep}^{nR_1}}
       +|{\cal Y}|^{-{2n^2\tau}{\ep}^{nR_2}} 
\\
& &
 +|{\cal X}|^{-2n^2 \tau{\ep}^{nR_1}}\cdot
  |{\cal Y}|^{-2n^2 \tau{\ep}^{nR_2}}
\\
&\leq& 3|{\cal X}|^{-2n^2 \tau{\ep}^{nR_1}}\cdot
        |{\cal Y}|^{-2n^2 \tau{\ep}^{nR_2}}.
\eeqno
which implies that for each fixed $\tau>0$, 
$\nu_{n,\tau}(R_1,$ $R_2,$ $|{\cal X}|,$ $|{\cal Y}|)$ 
decays double exponentially as $n$ tends to infinity. 
\end{pro}

From this {proposition}, we obtain the following corollary.   
\begin{co} \label{co:co0z}{\rm 
For any sequence of ID codes $\{(n,$ $N_1,$, $N_2,$ 
$\bar{\mu}_n,$ $\bar{\lambda}_n)$
$\}_{n=1}^{\infty}$ satisfying $\bar{\mu}_n+$ $\bar{\lambda}_n$ $ < 1$, 
$n=1,2,\cdots$, if 
$$
\liminf_{n\to\infty} r_{i,n} \geq R_i,\: i=1,2, 
$$
then, for any $\delta>0$, there exists $n_0=n_0(\delta)$ 
such that for $n\geq n_0$, 
\beqa
1-\bar{\mu}_n-\bar{\lambda}_n 
&\leq & \Omega_{n,\gamma}(R_1-\delta, R_2-\delta\Vl W^n)
\nonumber\\
& & + \nu_{n,\tau}(R_1-\delta,R_2-\delta,|{\cal X}|,|{\cal Y}|)\,. 
\label{eqn:ieqco0z}
\eeqa
}
\end{co} 

It immediately follows from Theorem A, Corollary \ref{co:co0z} 
and Property \ref{pr:pr1} part b) that the following strong 
converse theorem holds.  

\begin{Th}\label{th:coHVzz}
For any sequence of ID codes $\{(n, N_1,$ $N_2,$ 
$\bar{\mu}_n,$ $\bar{\lambda}_n)$
$\}_{n=1}^{\infty}$ satisfying $\bar{\mu}_n$ $+$ $\bar{\lambda}_n$ $<1$, 
$n=1,2,\cdots$, if  
\beqno
\liminf_{n\to\infty} r_{i,n} \geq R_i, i=1,2, \quad  
(R_1,R_2) \notin  \overline{\cal C}^{\prime}({\vc W})\,,
\eeqno
then, 
$$
\liminf_{n\to\infty}\{\bar{\mu}_n+\bar{\lambda}_n\}=1, 
$$
which implies that for any $\mu\geq 0, \lambda\geq 0,$
$\mu + \lambda < 1$ and any noisy channel ${\vc W}$, 
$$
\underline{\cal C}({\vc W})\subseteq 
{\cal C}_{\rm ID}(\mu,\lambda|{\vc W})
\subseteq 
\cida(\mu,\lambda|{\vc W})
\subseteq 
\overline{\cal C}^{\prime}({\vc W})\,.
$$
In particular, if 
$$ 
\underline{\cal C}({\vc W})
=\overline{\cal C}({\vc W})
=\overline{\cal C}^{\prime}({\vc W}),
$$ 
then, for any $\mu\geq 0,\lambda \geq 0, $ $\mu+\lambda < 1$, 
$$
\underline{\cal C}({\vc W})=
{\cal C}_{\rm ID}(\mu,\lambda|{\vc W})
=\cida(\mu,\lambda|{\vc W})
=\overline{\cal C}({\vc W})
=\overline{\cal C}^{\prime}({\vc W}).
$$
Furthermore, $\bar{\mu}_n+\bar{\lambda}_n$ converges to one as $n\to\infty$ 
at rates above the ID capacity. This implies that 
the strong converse property holds with respect to the sum of 
two types of error probabilities.
\end{Th}

\section{Proof of Results}

In this section we shall give the proofs of the results 
stated in the previous section.  

For the proofs of {Propositions \ref{pro:thM2} and \ref{pro:thM3}}, we
first formulate a new resolvability problem for the general MAC, that
is, an approximation problem of output random variables via MACs.  We
consider this problem and derive an upper bound of the approximation
error. This upper bound is useful for analyzing the error probability
of identification outside the ID capacity region.  Next, we prove
Propositions \ref{pro:thM2} and \ref{pro:thM3} based on a new method
of converting the direct coding theorem for the MAC resolvability
problem into the converse coding theorem of the ID via MACs.  Han and
Verd\'u \cite{hv2} provided a method of converting the direct coding
theorem for the channel resolvability problem into the converse coding
theorem of the ID channel.  Our method is an extension of their method
in the case of MACs.

\subsection{MAC Resolvability Problem}

\begin{df}
{\rm 
Let $U_{M_i}, i=1,2$ 
be the uniform random variables taking values in 
${\cal U}_{M_1}=$
$\{1,2,$ $\cdots,$ $M_i\}$. 
By two maps 
$\resenc_1 :$ ${\cal U}_{M_1}\to$ ${\cal X}^n$ and 
$\resenc_2 :$ ${\cal U}_{M_2}\to$ ${\cal Y}^n$, 
the uniform random variables $U_{M_1}$ and $U_{M_2}$ 
is transformed into the random variable 
$\tilde{X}^n=\resenc_1(U_{M_1})$ and 
$\tilde{Y}^n=\resenc_2(U_{M_2})$, respectively.
Let ${\cal P}_{M_1}$ $({\cal X}^n)$ 
and ${\cal P}_{M_2}$ $({\cal Y}^n)$
be sets of all probability
distributions on ${\cal X}^n$ that can be created by the 
transformation of $U_{M_1}$ and $U_{M_2}$. 
Elements of ${\cal P}_{M_1}({\cal X}^n)$ 
and ${\cal P}_{M_2}({\cal Y}^n)$, respectively  
are called $M_1$ and $M_2$-types.
Every random variable $\tilde{X}^n=\resenc_1($ $U_{M_1}$ $)$ 
created by some transformation map 
$\resenc_1:$ ${\cal U}_{M_n}$ $\to$ 
${\cal X}^n$ and $U_{M_1}$ has $M_1$-type.
Similarly, every random variable $\tilde{Y}^n=\resenc_2($ $U_{M_2}$ $)$ 
created by some transformation 
map $\resenc_2:$ ${\cal U}_{M_2}$ $\to$ 
${\cal Y}^n$ and $U_{M_2}$ has $M_2$-type.}
\end{df}

\begin{df}\label{df:df7}{\rm 
For 
$\resenc_1:{\cal U}_{M_1}$ $\to {\cal X}^n$ 
and  
$\resenc_2:{\cal U}_{M_2}$ $\to {\cal Y}^n$, 
define ${P}_{\tilde{X}^n}=$ $P_{\resenc_1(U_{M_1})}$ and  
       ${P}_{\tilde{Y}^n}=$ $P_{\resenc_2(U_{M_2})}$.
We use ${P}_{\tilde{X}^n}$ and ${P}_{\tilde{Y}^n}$
as approximations of $X^n$ and $Y^n$, respectively. 
Let 
$\tilde{Q}^{(1)}$ be a response of $(P_{\tilde{X}^n}$,$P_{{Y}^n})$ 
and let
$\tilde{Q}^{(2)}$ be a response of $(P_{{X}^n}$,$P_{\tilde{Y}^n})$. 
Let $\tilde{Q}^{(3)}$ 
be a response of $(P_{\tilde{X}^n}$,$P_{\tilde{Y}^n})$. 
Set 
$$
\underline{\tilde{Q}}\defeq
(\tilde{Q}^{(1)}, \tilde{Q}^{(2)}, \tilde{Q}^{(3)}).
$$
Let $\tilde{\cal Q}^{(t)}$, $t=1,2,3$, be sets 
of all responses $\tilde{Q}^{(t)}$.
} 
\end{df}

The following is a lemma on the cardinalities of  
${\cal P}_{M_1}($ ${\cal X}^n),$ ${\cal P}_{M_2}($ ${\cal Y}^n)$
and $\tilde{\cal Q}^{(t)}$, $t=1,2,3$. 

\begin{lm}{\label{lm:countlm}\rm
$\quad$
\begin{itemize}
\item[a)]
\beqno
|{\cal P}_{M_1}({\cal X}^n)| 
&\leq & |{\cal X}|^{nM_1},
|{\cal P}_{M_2}({\cal Y}^n)|
\leq |{\cal Y}|^{nM_2}.
\eeqno
\item[b)]
\beqno
&& |\tilde{\cal Q}^{(1)}| \leq |{\cal P}_{M_1}({\cal X}^n)|,
   |\tilde{\cal Q}^{(2)}| \leq |{\cal P}_{M_2}({\cal Y}^n) |\,,
\\ 
&& |\tilde{\cal Q}^{(3)}|
   \leq |{\cal P}_{M_1}({\cal X}^n)||{\cal P}_{M_2}({\cal Y}^n )|.
\eeqno
\end{itemize}
}
\end{lm}

Now we use $\underline{\tilde{Q}}$ 
as an approximation of $Q$. In this case we are interested in the
asymptotic behavior of the following triple of approximation errors 
$$
(d(Q,\tilde{Q}^{(1)}), d(Q,\tilde{Q}^{(2)}), d(Q,\tilde{Q}^{(3)}))
$$ 
measured by the variational distance. We shall derive 
explicit upper bounds of $d(Q, \tilde{Q}^{(t)}), t=1,2,3$.
This result is a mathematical core of the converse coding theorem 
for the ID via MACs. 
\begin{lm}\label{lm:mlm}{\rm Set
$M_t= \lceil{\ep}^{nR_t}\rceil$, $t=1,2$, 
where $\lceil a \rceil$ is the minimum integer not below $a$.
Let $S_i,i=1,2,3$ be arbitrary prescribed subsets 
of ${\cal X}^n$ $\times {\cal Y}^n$ $\times {\cal Z}^n$.
Let ($X^n, Y^n)$ be a pair of two independent random variables 
with distribution $(P_{X^n},$ $P_{Y^n})$. Let $Q$ be a response of 
$(P_{X^n},$ $P_{Y^n})$. Then, for any 
$(P_{X^n},$ $P_{Y^n})$ and its response $Q$, 
there exist 
$\resenc_1:{\cal U}_{M_1}$ $\to {\cal X}^n$ 
and 
$\resenc_2:{\cal U}_{M_2}$ $\to {\cal Y}^n$ 
such that the three variational distances 
$d(Q,\tilde{Q}^{(t)}),$ $t=1,2,3$
satisfies the following:
\beqno
&&d(Q,\tilde{Q}^{(t)})
\nonumber\\
&\leq& 4\Av\left[{\vc 1}_{S_t^c}(X^n,Y^n,Z^n)\right]
+3\sqrt{\zeta_{n,t,S_t}},\mbox{ for }t=1,2,3. 
\eeqno
}
\end{lm}

The proof of the above lemma is given in the appendix. 

\subsection{Proofs of {Propositions and Corollaries}}

In this subsection we { prove }
Propositions \ref{pro:thM2} and \ref{pro:thM3}
and Corollaries \ref{co:co0} and \ref{co:co0z} 
stated in the previous section. We first prove 
Propositions \ref{pro:thM2} and \ref{pro:thM3} 
using Lemmas \ref{lm:countlm} and \ref{lm:mlm}. 
Next we prove Corollaries 
\ref{co:co0} and \ref{co:co0z} respectively, 
using Propositions \ref{pro:thM2} and \ref{pro:thM3}. 

{\it Proof of Proposition \ref{pro:thM2}: }
Let $P_{X^n|i}\in {\cal P}({\cal X}^n), i\in {\cal N}_1$ 
and $P_{Y^n|j}\in {\cal P}({\cal Y}^n), j\in {\cal N}_2$, 
be codewords of $(n,N_1,N_2,\mu_n,\lambda_n)$ code of 
the ID channel and $D_{i,j} \subseteq {\cal Z}^n,$ $i\in {\cal N}_1 $, 
$j\in {\cal N}_2 $ be decoding regions corresponding 
to the codewords. Let the 
response $P_{X^n|i}P_{Y^n|i}W^n$ of $(P_{X^n|i},$ $P_{Y^n|j})$ 
be denoted by
$Q_{ij}$. We choose $S_i=T_{i,\gamma}$, $i=1,2,3$. 
Then, by Lemma \ref{lm:mlm}, there exists 
$ \underline{\tilde{Q}}$ such that
\beq
d(Q_{ij},\tilde{Q}^{(t)})\leq \eta_{n,t}(P_{X^n}, P_{Y^n}), t=1,2,3,
\label{eqnzzz}
\eeq
where we put 
\beqno
\eta_{n,t} 
(P_{X^n}, P_{Y^n})
& \defeq & \Omega_{n,t, \gamma}(R_t,P_{X^n}, P_{Y^n}|W^n), t=1,2,
\\
\eta_{n,3}(P_{X^n},P_{Y^n})
& \defeq & \Omega_{n,3,\gamma}(R_1,R_2,P_{X^n},P_{Y^n}|W^n).
\eeqno 
For simplicity of notation we set
$
\eta_n\defeq \Omega_{n,\gamma}(R_1,R_2|W^n).
$  
Then by the definition of $\Omega_{n,\gamma}(R_1,R_2|W^n)$, 
we have
\beq
\eta_n=\sup_{
\scs (P_{X^n},P_{Y^n})
         \atop{\scs \in {\cal P}({\cal X}^n)\times {\cal P}({\cal Y}^n)}  
         }
\min_{t=1,2,3}\{\eta_{n,t}(P_{X^n},P_{Y^n})\}.
\label{eqnzzz1}
\eeq 
From (\ref{eqnzzz}) and (\ref{eqnzzz1}), it follows that
for any $Q_{ij}$, there exists $t\in\{1,2,3\}$ 
and $\tilde{Q}^{(t)}$ $\in \tilde{\cal Q}^{(t)}$ 
such that
$
d(Q_{ij},\tilde{Q}^{(t)})
\leq \eta_{n}.
$ 
Define 
\beqno
{\cal L}^{(t)}
 &\defeq& \{(i,j)\in {\cal N}_1\times {\cal N}_2: 
\\
& &\qquad d(Q_{ij},\tilde{Q}^{(t)})\leq \eta_{n}
\mbox{ for some } \tilde{Q}^{(t)}
\in \tilde{\cal Q}^{(t)}
\}.
\eeqno
Since 
$$
{\cal L}^{(1)}\cup {\cal L}^{(2)}\cup {\cal L}^{(3)}
={\cal N}_1\times {\cal N}_2\,,
$$
we have  
\beq
|{\cal L}^{(t)}|\geq \frac{1}{3}N_1N_2
\mbox{ for some }t\in\{1,2,3\}. 
\label{eqn:prth00a}
\eeq
Set $a_t\defeq |\tilde{\cal Q}^{(t)}|$, $t=1,2,3$. Note that 
$M_t\leq 2\ep^{nR_t},$ $t=1,2$. Then 
by Lemma \ref{lm:countlm}, we have 
\beqno
a_1\leq |{\cal X}|^{2\ep^{nR_1}},
a_2\leq |{\cal Y}|^{2\ep^{nR_2}},
a_3\leq |{\cal X}|^{2\ep^{nR_1}}
\cdot |{\cal Y}|^{2\ep^{nR_2}}. 
\eeqno
Set
\beqno
b_1\defeq |{\cal X}|^{2\ep^{nR_1}},
b_2\defeq |{\cal Y}|^{2\ep^{nR_2}},
b_3 \defeq |{\cal X}|^{2\ep^{nR_1}}\cdot |{\cal Y}|^{2\ep^{nR_2}}. 
\eeqno
Now, we suppose that the inequality (\ref{eqn:prth00a}) 
holds for $t=1$. Set
$$
{\cal L}^{(1)}_{1|2}(j)\defeq\{i:(i,j)\in {\cal L}^{(1)}\}.
$$
Then, we have 
\beqno
|{\cal L}^{(1)}_{1|2}(j)|\geq \frac{1}{3}N_1
\mbox{ for some }j. 
\eeqno
Then if 
$$
\frac{1}{3}N_1 \geq 3^{2\ep^{nR_1}-1}\cdot b_1
\geq 3b_1 > a_1 = |\tilde{\cal Q}^{(1)}|
$$
or equivalent to 
\beqno
r_{1,n}
&\geq& R_1+\frac{\log n}{n} +\frac{1}{n}\log\log(3|{\cal X}|)^2,
\eeqno
there exist two pairs $(i,j)$ and $(k,j)$,  
$i\ne k$ and $\tilde{Q}^{(1)}$ $\in \tilde{\cal Q}^{(1)}$ 
such that 
$$ 
d(Q_{ij},\tilde{Q}^{(1)})\leq \eta_n\,, 
d(Q_{kj},\tilde{Q}^{(1)})\leq \eta_n\,. 
$$
For the above two pairs, we have
\beq
d(Q_{ij},Q_{kj})
\leq d(Q_{ij},\tilde{Q}^{(1)})+d(Q_{kj},\tilde{Q}^{(1)}) 
\leq  2\eta_n.
\label{eqn:prth01Z}
\eeq
On the other hand, we have  
\beqno
d(Q_{ij},Q_{kj}) 
&\geq & 2\left[Q_{ij}(D_{i,j})-Q_{kj}(D_{i,j})\right]
\nonumber\\
&\geq & 2\left(1-\mu_n-\lambda_n\right)\,,
\eeqno
which together with (\ref{eqn:prth01Z}) yields that 
$1-\mu_n-\lambda_n $ $\leq \eta_n.$
Next, we suppose that the inequality (\ref{eqn:prth00a}) 
holds for $t=2$. By an argument quite similar to 
the previous one, we can prove that  
if 
$$
\frac{1}{3}N_2 \geq 3^{2\ep^{nR_2}-1}\cdot b_2 
\geq 3b_2 >a_2 = |\tilde{\cal Q}^{(2)}|
$$
or equivalent to 
\beqno
r_{2,n}
&\geq& R_2+\frac{\log n}{n}+\frac{1}{n} \log \log (3|{\cal Y}|)^2,
\eeqno
we have $1-\mu_n-\lambda_n $ $\leq \eta_n$.
Finally, we suppose that the inequality (\ref{eqn:prth00a}) 
holds for $t=3$. 
Since 
$$
\frac{1}{3}N_1N_2 \geq 
3^{2(\ep^{nR_1} + \ep^{nR_2})-1}\cdot b_1b_2\geq 3b_1b_2 
> a_1a_2\geq |\tilde{\cal Q}^{(3)}|
$$
there exist two pairs $(i,j)$ and $(k,l)$, 
$(i,j)\ne (k,l)$ and $\tilde{Q}^{(3)}$ $\in \tilde{\cal Q}^{(3)}$ 
such that 
$$ 
d(Q_{ij},\tilde{Q}^{(3)})\leq \eta_n, 
d(Q_{kl},\tilde{Q}^{(3)})\leq \eta_n. 
$$
For the above two pairs, we have
\beq
d(Q_{ij},Q_{kl})
\leq d(Q_{ij},\tilde{Q}^{(3)})+d(Q_{kl},\tilde{Q}^{(3)}) 
\leq  2\eta_n.
\label{eqn:prth01Za}
\eeq
On the other hand, we have  
\beqno
d(Q_{ij},Q_{kl}) 
&\geq & 2\left[Q_{ij}(D_{i,j})-Q_{kl}(D_{i,j})\right]
\nonumber\\
&\geq & 2\left(1-\mu_n-\lambda_n\right)\,,
\eeqno
which together with (\ref{eqn:prth01Za}) yields that 
$1-\mu_n-\lambda_n $ $\leq \eta_n.$
This completes the proof of Proposition \ref{pro:thM2}. 
\hfill\IEEEQED

{\it Proof of Proposition \ref{pro:thM3}:}
Let $P_{X^n|i}\in {\cal P}({\cal X}^n), i\in {\cal N}_1$, and
$P_{Y^n|j}\in {\cal P}({\cal Y}^n), j\in {\cal N}_2$, be codewords of
$(n,N_1,N_2,\bar{\mu}_n,\bar{\lambda}_n)$ code of the ID channel and
$D_{ij} \subseteq {\cal Z}^n,$ $i\in {\cal N}_1 $, $j\in {\cal N}_2 $
be decoding regions corresponding to the codewords.  Let the response
$P_{X^n|i}P_{Y^n|i}W^n$ of $(P_{X^n|i},$ $P_{Y^n|j})$ be denoted by
$Q_{ij}$.  
For 
$\tilde{Q}^{(t)}\in$ $\tilde{\cal Q}^{(t)}$, 
$t=1,2,3$, 
define 
$$
{\cal S}_t(\tilde{Q}^{(t)})
\defeq
\left\{(i,j)\in {\cal N}_1\times {\cal N}_2 : 
d(Q_{ij},\tilde{Q}^{(t)})\leq \eta_n\right\}.  
$$ 
For $t=1,2,3$, set
\beqno
&&\tilde{\cal Q}_0^{(t)}
\defeq
\left\{
\tilde{Q}^{(t)} \in \tilde{\cal Q}^{(t)}: 
|{\cal S}_t(\tilde{Q}^{(t)})|\geq 1
\right\}.
\eeqno
Then, the validity of Lemma \ref{lm:mlm} implies that
\beqno
& &{\cal L}^{(t)}=\bigcup_{\tilde{Q}^{(t)}\in \tilde{\cal Q}_0^{(t)}} 
{\cal S}_t(\tilde{Q}^{(t)})\mbox{ for }t=1,2,3,
\\
& &\bigcup_{t=1}^3\bigcup_{\tilde{Q}^{(t)}\in \tilde{\cal Q}_0^{(t)}} 
{\cal S}_t(\tilde{Q}^{(t)})={\cal N}_1\times {\cal N}_2.
\eeqno
Define
\beqno
\tilde{\cal Q}_{1}^{(1)}&\defeq&
\Bigl\{ \tilde{Q}^{(1)}\in \tilde{\cal Q}^{(1)}: \Bigr. 
\ba[t]{l} 
{\cal S}_1(\tilde{Q}^{(1)})\mbox{ consists of pairs }(i,j)\\
\mbox{ such that for fixed }j \mbox{ we have }\\
\mbox{ only one index }i \Bigl.\Bigr\},
\ea 
\\
\tilde{\cal Q}_{2}^{(1)}&\defeq&
\Bigl\{ \tilde{Q}^{(1)}\in \tilde{\cal Q}^{(1)}: \Bigr. 
\ba[t]{l} 
{\cal S}_1(\tilde{Q}^{(1)})\mbox{ consists of pairs }(i,j)\\
\mbox{ such that for fixed }j \mbox{ we have }\\
\mbox{ more than two indexes }i \Bigl.\Bigr\},
\ea
\\
\tilde{\cal Q}_{1}^{(2)}&\defeq&
\Bigl\{ \tilde{Q}^{(2)}\in \tilde{\cal Q}^{(2)}: \Bigr. 
\ba[t]{l} 
{\cal S}_2(\tilde{Q}^{(2)})\mbox{ consists of pairs }(i,j)\\
\mbox{ such that for fixed }i \mbox{ we have }\\
\mbox{ only one index }j \Bigl.\Bigr\},
\ea 
\\
\tilde{\cal Q}_{2}^{(2)}&\defeq&
\Bigl\{ \tilde{Q}^{(2)}\in \tilde{\cal Q}^{(2)}: \Bigr. 
\ba[t]{l} 
{\cal S}_2(\tilde{Q}^{(2)})\mbox{ consists of pairs }(i,j)\\
\mbox{ such that for fixed }i \mbox{ we have }\\
\mbox{ more than two indexes }j \Bigl.\Bigr\},
\ea
\\
\tilde{\cal Q}_1^{(3)}&\defeq&
\Bigl\{ \tilde{Q}^{(3)}\in \tilde{\cal Q}^{(3)}: \Bigr. 
\ba[t]{l} 
{\cal S}_3(\tilde{Q}^{(3)})\mbox{ consists of pairs }(i,j)\\
\mbox{ with one index pair }(i,j) \Bigl.\Bigr\},
\ea 
\\
\tilde{\cal Q}_{2}^{(3)}&\defeq&
\Bigl\{ \tilde{Q}^{(3)} \in \tilde{\cal Q}^{(3)}: \Bigr. 
\ba[t]{l} 
{\cal S}_3(\tilde{Q}^{(3)})\mbox{ consists of more }\\
\mbox{ than two index pairs }(i,j) \Bigl. \Bigr\}.
\ea
\eeqno
It is obvious that
\beqno
\tilde{\cal Q}_1^{(t)} \cup \tilde{\cal Q}_2^{(t)} 
 = \tilde{\cal Q}_{0}^{(t)},\mbox{ }t=1,2,3.
\eeqno
Observe that if $\tilde{Q}^{(1)} \in\tilde{\cal Q}_{2}^{(1)}$, 
for any $(i,j)\in {\cal S}_1(\tilde{Q})$, 
there exists an index $k\ne i$ 
such that $(k,j) \in{\cal S}_1(\tilde{Q})$. 
Then, we have 
\beqa
& & 1-\mu_{n,ij}- \lambda_{n,ij} 
\nonumber\\
&\leq & \left[Q_{ij}(D_{i,j})-Q_{kj}(D_{k,j})\right]
\leq  (1/2)d(Q_{ij}, Q_{kj})
\nonumber\\
&\leq & (1/2)\left[d(Q_{ij},\tilde{Q}^{(1)}) 
                  + d(Q_{kj},\tilde{Q}^{(1)})\right]
\leq \eta_n.
\label{eqn:prth1aa}
\eeqa
Similarly, if $\tilde{Q}^{(2)}\in\tilde{\cal Q}_{2}^{(2)}$, 
for any $(i,j)\in {\cal S}_2(\tilde{Q}^{(2)})$, 
there exists an index $l\ne j$ 
such that $(i,l) \in{\cal S}_2(\tilde{Q}^{(2)})$. 
Then, we have 
\beqa
1-\mu_{n,ij}- \lambda_{n,ij}
\leq \eta_n.
\label{eqn:prth1aaz}
\eeqa
If $\tilde{Q}^{(3)} \in\tilde{\cal Q}_2^{(3)}$, 
for any $(i,j)\in {\cal S}_3(\tilde{ Q}^{(3)})$, 
there exists an index $(i,j)\ne (k,l)$ 
such that $(k,l) \in{\cal S}_3(\tilde{Q}^{(3)})$. 
Then, we have 
\beqa
1-\mu_{n,ij}- \lambda_{n,ij} 
\leq \eta_n.
\label{eqn:prth1aazz}
\eeqa
We obtain the following chain of inequalities: 
\beqa
&&
1-\bar{\mu}_n-\bar{\lambda}_n
\nonumber\\
&=&\frac{1}{N_1N_2}
       \sum_{(i,j)\in {\cal N}_1 \times {\cal N}_2 }
         \hspace*{-3mm}(1- \mu_{n,ij}-\lambda_{n,ij})
\nonumber\\
&\leq &
\frac{1}{N_1N_2}
       \sum_{t=1}^3\sum_{(i,j)\in {\cal L}^{(t)}}
         (1- \mu_{n,ij}-\lambda_{n,ij})
\nonumber\\
&=&\frac{1}{N_1N_2}
\sum_{t=1}^3\sum_{\tilde{Q}^{(t)}
\in\tilde{\cal Q}_0^{(t)}} 
\sum_{(i,j) 
\in{\cal S}_t(\tilde{Q}^{(t)})}
    \hspace*{-3mm}(1- \mu_{n,ij}-\lambda_{n,ij})
\nonumber\\
&=&\frac{1}{N_1N_2}
   \sum_{t=1}^3\sum_{\tilde{Q}^{(t)}\in\tilde{\cal Q}_{1}^{(t)} }
   \sum_{(i,j) \in{\cal S}_t(\tilde{Q}^{(t)})}
    \hspace*{-3mm}(1-\mu_{n,ij}-\lambda_{n,ij})                
\nonumber\\
&&+\frac{1}{N_1N_2}
  \sum_{t=1}^3\sum_{\tilde{Q}^{(t)}\in\tilde{\cal Q}_{2}^{(t)} }
  \sum_{(i,j)\in{\cal S}_t(\tilde{Q}^{(t)})}
   \hspace*{-3mm}(1- \mu_{n,ij}-\lambda_{n,ij})                               
\nonumber\\
&\MLeq{a}& 
\sum_{t=1}^2\frac{|\tilde{\cal Q}_{1}^{(t)} |}{N_t} 
+\frac{|\tilde{\cal Q}_{1}^{(3)}|}{N_1N_2} 
+\eta_n
\nonumber\\
&\leq& \frac{|{\cal P}_{M_1}({\cal X}^n)| }{N_1} 
       +\frac{|{\cal P}_{M_2}({\cal Y}^n)| }{N_2} 
\nonumber\\
&  &   \qquad\qquad+\frac{|{\cal P}_{M_1}({\cal X}^n)||{\cal P}_{M_2}({\cal Y}^n)|}
        {N_1N_2}+\eta_n
\nonumber\\
&\MLeq{b}& 
         \frac{|{\cal X}|^{2n\ep^{nR_1}}}{N_1}
         +\frac{|{\cal Y}|^{2n\ep^{nR_2}}}{N_2}
\nonumber\\
& &    \qquad\qquad +\frac{|{\cal X}|^{2n\ep^{nR_1}}|{\cal Y}|^{2n\ep^{nR_2}}}{N_1N_2}
       +\eta_n.
\label{eqn:avieq2a}
\eeqa
Step (a) follows from 
 (\ref{eqn:prth1aa})
-(\ref{eqn:prth1aazz}).  
Step (b) follows from Lemma \ref{lm:countlm} and 
$M_t\leq 2\ep^{nR_t}, t=1,2.$ Then, if 
    $N_1\geq $ $|{\cal X}|^{2n\ep^{n(R_1+\tau)}}$ 
and $N_2\geq $ $|{\cal Y}|^{2n\ep^{n(R_2+\tau)}}$ or equivalent to
\beqno
r_{1,n} &\geq& R_1+\tau +\frac{\log n}{n} 
              +\frac{1}{n}\log\log |{\cal X}|^2\,,
\nonumber\\
r_{2,n} &\geq& R_2+\tau +\frac{\log n}{n}
              +\frac{1}{n}\log\log|{\cal Y}|^2\,,
\eeqno
from (\ref{eqn:avieq2a}), we have 
\beqno
&& 1-\bar{\mu}_n-\bar{\lambda}_n
\nonumber\\
&\leq& \frac{|{\cal X}|^{2n\ep^{nR_1}}}
            {|{\cal X}|^{2n\ep^{n(R_1+\tau)}}}
      +\frac{|{\cal Y}|^{2n\ep^{nR_2}}}
            {|{\cal Y}|^{2n\ep^{n(R_2+\tau)}}}
\nonumber\\ 
& &
+\frac{|{\cal X}|^{2n\ep^{nR_1}}|{\cal Y}|^{2n\ep^{nR_2}}}
{|{\cal X}|^{2n\ep^{n(R_1+\tau)}}
|{\cal Y}|^{2n\ep^{n(R_2+\tau)}}
}
+\eta_n
\nonumber\\
&= &\nu_{n,\tau}(R_1,R_2,|{\cal X}|,|{\cal Y}|)
    +\Omega_{n,\gamma}(R_1,R_2,W^n)\,.  
\eeqno
This completes the proof of Proposition \ref{pro:thM3}.
\hfill\IEEEQED

{\it Proof of Corollary \ref{co:co0}:}
We assume that a sequence of ID codes
$\{(n, N_1, N_2,$ $\mu_n,$ $\lambda_n)$ $\}_{n=1}^{\infty}$ satisfies 
$\mu_n +\lambda_n < 1,n=1,2,\cdots,$ and 
\beq
\ds\liminf_{n\to\infty}\frac{1}{n}\log\log N_i \geq R_i, i=1,2.
\label{eqn:ratelim}
\eeq
Since
\beqno 
\lim_{n\to \infty}\left[ \frac{\log n}{n}
+\frac{1}{n}\log \log \left(3|{\cal X}|\right)^2\right]=0\,,
\label{eqn:rcndddx}  
\\
\lim_{n\to \infty}\left[ \frac{\log n}{n}
+\frac{1}{n}\log \log \left(3|{\cal Y}|\right)^2\right]=0\,,
\label{eqn:rcndddy}  
\eeqno
there exists $n_1=n_1(\delta,|{\cal X}|,|{\cal Y}|)$ such that
for any $n\geq n_1$   
\beqno 
\frac{\log n}{n}
+\frac{1}{n}\log\log \left(3|{\cal X}|\right)^2\leq \frac{\delta}{2},
\\
\frac{\log n}{n}
+\frac{1}{n}\log\log\left(3|{\cal Y}|\right)^2\leq \frac{\delta}{2}.
\\
\eeqno
On the other hand, by virtue of (\ref{eqn:ratelim}), there exists 
$n_2=n_2(\delta)$ such that for any $n \geq n_2$
$$
\frac{1}{n}\log\log N_i \geq R_i-\frac{\delta}{2}, i=1,2. 
$$
Set $n_0=n_0(\delta,|{\cal X}|)=\max\{n_1,n_2\}.$ Then, 
for any $n \geq n_0$, we have
\beqno 
& &\frac{1}{n}\log\log N_1 \geq R_1-\delta 
   +\frac{\log n}{n}+\frac{1}{n}\log\log\left(3|{\cal X}|\right)^2,  
\\
& &\frac{1}{n}\log\log N_2 \geq R_2-\delta 
   +\frac{\log n}{n}+\frac{1}{n}\log\log\left(3|{\cal Y}|\right)^2. 
\eeqno
Applying {Proposition \ref{pro:thM2}} 
with respect to $R_i-\delta,i=1,2$, for $n\geq n_0$, 
we have (\ref{eqn:ieqco00}) of Corollary \ref{co:co0}.
\hfill\IEEEQED%

{\it Proof of Corollary \ref{co:co0z}:}
We assume that a sequence of ID codes
$\{(n, N_1, N_2,$ $\bar{\mu}_n,$ $\bar{\lambda}_n)$ $\}_{n=1}^{\infty}$ satisfies 
$\bar{\mu}_n +\bar{\lambda}_n < 1,n=1,2,\cdots,$ and 
\beq
\ds\liminf_{n\to\infty}\frac{1}{n}\log\log N_i \geq R_i, i=1,2.
\label{eqn:ratelimzz}
\eeq
We choose $\tau=(1/3)\delta$. Since
\beqno 
\lim_{n\to \infty}\left[ \frac{\log n}{n}
+\frac{1}{n}\log \log |{\cal X}|^2\right]=0,
\label{eqn:rcnx}  
\\
\lim_{n\to \infty}\left[ \frac{\log n}{n}
+\frac{1}{n}\log \log |{\cal Y}|^2\right]=0,
\label{eqn:rcny}  
\eeqno
there exists $n_1=n_1(\delta,|{\cal X}|,|{\cal Y}|)$ such that
for any $n\geq n_1$   
\beqno 
\tau + \frac{\log n}{n}
+\frac{1}{n}\log\log |{\cal X}|^2\leq \frac{\delta}{2},
\\
\tau+ \frac{\log n}{n}
+\frac{1}{n}\log\log |{\cal Y}|^2\leq \frac{\delta}{2}.
\\
\eeqno
On the other hand, by virtue of (\ref{eqn:ratelimzz}), 
there exists $n_2=n_2(\delta)$ such that
for any $n \geq n_2$
$$
\frac{1}{n}\log\log N_i \geq R_i-\frac{\delta}{2}, i=1,2. 
$$
Set $n_0=n_0(\delta,|{\cal X}|)=\max\{n_1,n_2\}\,.$ Then, 
for any $n \geq n_0$, we have
\beqno 
& &\frac{1}{n}\log\log N_1 \geq R_1-\delta 
   +\tau+ \frac{\log n}{n}+\frac{1}{n}\log\log |{\cal X}|^2,  
\\
& &\frac{1}{n}\log\log N_2 \geq R_2-\delta 
   +\tau +\frac{\log n}{n}+\frac{1}{n}\log\log |{\cal Y}|^2.  
\eeqno
Applying Proposition \ref{pro:thM2} 
with respect to  $R_i-\delta,i=1,2$, for $n\geq n_0$, 
we have (\ref{eqn:ieqco0z}) of Corollary \ref{co:co0z}.
\hfill\IEEEQED%

\section*{\empty}
\appendix


\subsection{
Proof of Property 
\protect{\ref{pr:pr1}}
}

\newcommand{\Ept}{

{\it Proof of Property \ref{pr:pr1} part b): }
When $(R_1,R_2)$ is an inner point of 
$\underline{\cal C}^{\prime}({\vc W})$, 
by definition of $\underline{\cal C}^{\prime}({\vc W})$, 
for some $({\vc U},$ ${\vc X})\in $ ${\cal P}_{\kappa}^{*}$ 
and any $\gamma\geq 0$,  
\beq
(R_1-\gamma, R_2-\gamma) 
\in \underline{\cal C}_{{\lvc U}{\lvc X}}({\vc W})\,,
\eeq
which together with the definition of 
$\underline{I}({\vc X};$${\vc Y}_1 | {\vc U})$ 
and $\underline{I}({\vc U};$ ${\vc Y}_2),$ 
yields that for some $({\vc U},$ ${\vc X})\in $ ${\cal P}_{\kappa}^{*}$
and any $\gamma\geq 0$, 
\beqa
\Wid& &\lim_{n\to\infty}\Pr\left[\frac{1}{n}i_{U^nX^nY_1^n}(X^n;Y_1^n|U^n)
        \leq R_1-\gamma\right]=0\,,
\label{eqn:apdx02a}\\
\Wid& &\lim_{n\to\infty}\Pr\left[\frac{1}{n}i_{U^nY_2^n}(U^n;Y_2^n)
        \leq R_2-\gamma\right]=0\,. 
\label{eqn:apdx02b}
\eeqa
Hence, by (\ref{eqn:apdx02a}) and (\ref{eqn:apdx02b}), for some 
$({\vc U},$ ${\vc X})\in $ ${\cal P}_{\kappa}^{*}$, 
for any $\gamma \geq 0$ and for $t=1,2$ 
\beq
\lim_{n\to\infty}
\Pr\left\{(U^n,X^n,Y_1^n,Y_2^n)\in T_{t,\gamma}\right\}=0\,,
\eeq
or equivalent to
\beq
\lim_{n\to\infty}\Omega_{n,t,\gamma}^{(1)}
 (R_t,P_{U^nX^n},W^n) =1\,.
\label{eqn:apdx02kk}
\eeq
On the other hand, by Property \ref{pr:pr0} part a), 
for any $({\vc U},$ ${\vc X})\in $ ${\cal P}_{\kappa}^{*}$
and any $\gamma>0\,,$ and for $t=1,2$
\beq
\lim_{n\to\infty}\Omega_{n,t,\gamma}^{(2)}(R_t,P_{U^nX^n},W^n) =0\,.
\label{eqn:apdx022}
\eeq
Hence, by (\ref{eqn:apdx02kk}) and (\ref{eqn:apdx022}), 
for some $({\vc U},$ ${\vc X})\in $ ${\cal P}_{\kappa}^{*}$
and for any $\gamma >0$, 
\beq
\lim_{n\to\infty}\Omega_{n,\gamma}(R_1,R_2,P_{U^nX^n},W^n)=1\,.
\label{eqn:apdx0551}
\eeq
From (\ref{eqn:apdx0551}) 
and the definition of $\Omega_{n,\gamma}(R_1,$ $R_2,W^n)$, 
for any $ \gamma >0$, we have  
\beq
\lim_{n\to\infty}\Omega_{n,\gamma}(R_1,R_2,W^n)=1\,. 
\label{eqn:apdx060}
\eeq
Furthermore, from (\ref{eqn:apdx02kk}) 
\beq
\liminf_{n\to\infty}\Omega_{n,0}(R_1,R_2,W^n)\geq 1\,. 
\label{eqn:apdx06}
\eeq
}

{\it Proof of Property \ref{pr:pr1} part b): }
We assume that $(R_1,R_2)\notin \overline{\cal C}^{\prime}({\vc W})$.  
Then there exists small positive number $\gamma_0$ such 
that for any 
$0\leq \gamma $ 
$ \leq \gamma_0$, we have
$$
(R_1-\gamma,R_2-\gamma)\notin \overline{\cal C}^{\prime}({\vc W}).
$$
Then, by the definition of $\overline{\cal C}^{\prime}({\vc W})$, 
for any $({\vc X}$, ${\vc Y})$ 
$\in {\cal S}_{I}$, we have   
$$
(R_1-\gamma, R_2- \gamma) 
\notin \overline{\cal C}^{\prime}({\vc X},{\vc Y}|{\vc W})\,,
$$
or equivalent to
\beqa
& &  R_1-\gamma > \overline{I}({\vc X};{\vc Z}|{\vc Y}),
\label{eqn:abba0}\\ 
\mbox{or}& &R_2-\gamma > \overline{I}({\vc Y};{\vc Z}|{\vc X}),
\label{eqn:abba10}\\ 
\mbox{or}& &
\left\{
\ba{l}
R_1-\gamma > \overline{I}({\vc X};{\vc Z}),
R_2-\gamma > \overline{I}({\vc Y};{\vc Z}),
\\
R_1+R_2-2\gamma > \overline{I}({\vc X}{\vc Y};{\vc Z}).
\ea
\right.
\label{eqn:abba1} 
\eeqa
We first assume that (\ref{eqn:abba0}) holds. 
Then by the definition of 
$\overline{I}({\vc X};{\vc Z}|{\vc Y})$, 
for any $\gamma\in [0,\gamma_0)$,
\beq
\liminf_{n\to \infty}
\Omega_{n,1,\gamma}^{(1)}(R_1, P_{X^n},P_{Y^n}\Vl W^n)=0.
\label{eqn:apdx20}
\eeq
We choose $\tau$ so that $\tau=(1/2)($ $\gamma+\gamma_0)$.
Then by Property \ref{pr:pr0} part a), we have 
\beqa
\lefteqn{\hspace*{-5mm}
 \Omega_{n,t,\gamma}^{(2)}(R_1,P_{X^n},P_{Y^n}\Vl W^n)}
\nonumber\\
&\leq&{\ep}^{-n\tau}+\Omega_{n,t,\tau}^{(1)}
(R_t,P_{X^n},P_{Y^n}\Vl W^n)
\nonumber\\
& & -\Omega_{n,t,\gamma}^{(1)}(R_1,P_{X^n}, P_{Y^n}\Vl W^n).
\label{eqn:apdx23}
\eeqa
From (\ref{eqn:apdx20}) and (\ref{eqn:apdx23}),
for any $\gamma\in [0,\gamma_0)$, 
\beq
\liminf_{n\to \infty}
\Omega_{n,1,\gamma}(R_1,P_{X^n},P_{Y^n}\Vl W^n)=0.
\label{eqn:apdx22}
\eeq
Next, we suppose that (\ref{eqn:abba10}) holds. 
In a manner quite similar to the case of (\ref{eqn:abba0}), we obtain
\beq
\liminf_{n\to \infty}
\Omega_{n,2,\gamma}(R_2,P_{X^n},P_{Y^n}\Vl W^n)=0. 
\label{eqn:apdx22z}
\eeq
Finally, we assume that (\ref{eqn:abba1}) holds.
Observe that
\beqa
\hspace*{-4mm}&    &
\Omega_{n,3,\gamma}^{(1)}(R_1,R_2, P_{X^n},P_{Y^n}\Vl W^n)
\nonumber\\
\hspace*{-4mm}&\leq &
{\rm Pr}\left\{R_1-\gamma< \frac{1}{n}i_{X^nZ^n}(X^n;Z^n)\right\}
\nonumber\\
\hspace*{-4mm}&    &
+{\rm Pr}\left\{R_2-\gamma< \frac{1}{n}i_{Y^nZ^n}(Y^n;Z^n)\right\}
\nonumber\\
\hspace*{-4mm}&   &
+{\rm Pr}\left\{R_1+R_2-2 \gamma< \frac{1}{n}i_{X^nY^nZ^n}(X^nY^n;Z^n)\right\}.
\label{eqn:apdx29}
\eeqa
By (\ref{eqn:abba1}), (\ref{eqn:apdx29}), and the definitions of 
$\overline{I}({\vc X};{\vc Z})$, 
$\overline{I}({\vc Y};{\vc Z})$, and 
$\overline{I}({\vc X}{\vc Y};{\vc Z})$, 
for any $\gamma \in [0,\gamma_0)$, we have  
\beq
\lim_{n\to\infty}\Omega_{n,3,\gamma}^{(1)}
(R_1,R_2, P_{X^n},P_{Y^n} \Vl W^n)=0. 
\label{eqn:apdx30}
\eeq
We choose $\tau$ so that $\tau=(1/2)($ $\gamma+\gamma_0)$.
By Property \ref{pr:pr0} part a), we have 
\beqa
\lefteqn{\hspace*{-5mm}
 \Omega_{n,3,\gamma}^{(2)}(R_1,R_2,P_{X^n},P_{Y^n}\Vl W^n)}
\nonumber\\
&\leq& 3{\ep}^{-n\tau}
+\Omega_{n,3,\tau}^{(1)}(R_1,R_2,P_{X^n},P_{Y^n}\Vl W^n)
\nonumber\\
& & -\Omega_{n,3,\gamma}^{(1)}(R_1,R_2, P_{X^n}, P_{Y^n}\Vl W^n).
\label{eqn:apdx31}
\eeqa
From (\ref{eqn:apdx30}) and (\ref{eqn:apdx31}),
for any $\gamma\in [0,\gamma_0)$, we have 
\beq
\lim_{n\to\infty}\Omega_{n,3,\gamma}
(R_1,R_2,P_{X^n},P_{X^n}\Vl W^n)=0. 
\label{eqn:apdx25}
\eeq
From (\ref{eqn:apdx22}), (\ref{eqn:apdx22z}), and (\ref{eqn:apdx25}),
we have 
$$
\lim_{n\to\infty}
\Omega_{n,\gamma}(R_1,R_2, P_{X^n},P_{Y^n}\Vl W^n)=0 
$$
for any $\gamma\in [0,\gamma_0)$ and 
for any  $({\vc X},$ $ {\vc Y})$ $\in {\cal S}_I$. 
Hence, by the definition of 
$\Omega_{n,\gamma}(R_1,$ $R_2\Vl W^n)$ 
we have for any $\gamma\in [0,\gamma_0)$,  
\beq
\lim_{n\to\infty}
\Omega_{n,\gamma}(R_1,R_2 \Vl W^n)=0, 
\label{eqn:apdx26}
\eeq
completing the proof. \hfill\IEEEQED

\subsection{Proof of Lemma \protect{\ref{lm:mlm}}}

In this appendix we shall prove Lemma \ref{lm:mlm}.
We first define several quantities necessary for the proof.

\begin{df}[Partial response(Steinberg \cite{stb})]{\rm
Let ($X^n,Y^n)$ be a pair of two independent random vectors with    
distribution $(P_{X^n},$ $P_{Y^n})$.
Let $S$ be a subset of 
${\cal X}^n\times$ 
${\cal Y}^n\times$ 
${\cal Z}^n$. 
Define a measure on ${\cal Z}^n$ by  
\beqa
Q_{S}({\vc z}) 
&=& \sum_{({\lvc x},{\lvc y})\in {\cal X}^n \times {\cal Y}^n }
W^n({\vc z}|{\vc x},{\vc y})
P_{{X}^n}({\vc x})
P_{{Y}^n}({\vc y})
\nonumber\\
&&\qquad \qquad\times {\vc 1}_S({\vc x},{\vc y},{\vc z}) 
\eeqa
We call the measure $Q_{S}$ 
{\it the partial response of }$(P_{{X}^n},$ $P_{{Y}^n})$ 
{\it on $S$ through noisy channel $W^n$}. 
By definition of the partial response, it is obvious that 
\beq
Q=Q_{S} + Q_{S^c}\,. 
\eeq
Note that $Q_S$ is no longer a probability measure.
}
\end{df}

%
%
Let $S_i, i=1,2,3$ be arbitrary subsets of 
${\cal X}^n \times$ 
${\cal Y}^n \times$ 
${\cal Z}^n$.
For $i=1,2,3$ define  
\beqno
{S}_{i,Z}&=&
\{
 {\vc z}\in {\cal Z}^n: ({\vc x},{\vc y},{\vc z})\in S_{i} 
\mbox{ for some }{\vc x},{\vc y}
\}\,, 
\\
{S}_{i,ZY}&=&
\{ ({\vc z},{\vc y})\in {\cal Z}^n \times {\cal Y}^n: 
 ({\vc x},{\vc y},{\vc z})\in S_{i}
\nonumber\\ 
 & & \mbox{ for some }{\vc x}\,\}\,.
\eeqno
For ${\vc z}\in S_{i,Z}$ define
\beqno
S_{i,XY|Z}({\vc z})&=&
\{
({\vc x},{\vc y})\in {\cal X}^n\times {\cal Y}^n: 
({\vc x},{\vc y},{\vc z})\in S_{i}
\}\,,  
\\
S_{i,Y|Z}({\vc z})&=&
\{
{\vc y}\in {\cal Y}^n: ({\vc z},{\vc y})\in S_{i,ZY}
\}\,.  
\eeqno
For $({\vc z},{\vc y})\in S_{i,ZY}$ define
$$
S_{i,X|ZY}({\vc z},{\vc y})=
\{
{\vc x}\in {\cal X}^n: ({\vc x}, {\vc y},{\vc z})\in S_{i}
\}\,.  
$$

{\it Proof of Lemma \ref{lm:mlm}:}
The proof consists of three steps.

{\it Step 1 (Random Coding Argument):}
Let $X_j^n, j\in {\cal U}_{M_1}$ be a sequence of independently and
identically distributed (i.i.d.) random variables each with
distribution $P_{X^n}\in {\cal P}({\cal X}^n)$. Each output of the
above random variables define a map $\resenc_1 :{\cal U}_{M_1}$ $\to
{\cal X}^n$. We use this randomly selected $\resenc_1$ as a
transformation map. 
Define      
$$
\chi_{{\lvc x}}({\vc x}^{\prime})=
\left\{\ba{ll}
1&\mbox{ if } {\vc x}={\vc x}^{\prime}\\
0&\mbox{ else }
\ea\right.
$$
Using the above $\resenc_i$, the input distribution 
$\tilde{P}_{X^n}=$ $\{\tilde{P}_{X^n}({\vc x})$ $\}_{{\vc x}\in {\cal X}^n}$ 
of $\resenc_1(U_{M_1})$ becomes a random variable, having the
form 
$$
\tilde{P}_{X^n}({\vc x})
=\tilde{P}_{[X_1^n,X_2^n,\cdots,X_{M_1}^n]}({\vc x})
=\frac{1}{M_1}
   \sum_{j=1}^{M_1}\chi_{\lvc x}(X_j^n)\,.
$$
Similarly, let $Y_j^n, j\in {\cal U}_{M_2}$ be a sequence of 
i.i.d. random variables each with distribution $P_{Y^n}\in {\cal P}({\cal
Y}^n)$. Each output of the above random variables define a map
${\resenc}_2 :{\cal U}_{M_2}$ $\to {\cal Y}^n$. We use this randomly
selected $\resenc_2$ as a transformation map.
Using the above $\resenc_2$, the input distribution 
$\tilde{P}_{Y^n}=$ $\{\tilde{P}_{Y^n}({\vc y})$ 
$\}_{{\vc y}\in {\cal Y}^n}$ 
of $\resenc_2(U_{M_2})$ becomes 
a random variable, having the
form 
$$
\tilde{P}_{Y^n}({\vc y})
=\tilde{P}_{[Y_1^n,Y_2^n,\cdots,Y_{M_2}^n]}({\vc y})
=\frac{1}{M_2}
   \sum_{j=1}^{M_2}\chi_{\lvc y}(Y_j^n)\,.
$$
Note that
\beqa
 {\bf E}\left[\tilde{Q}^{(1)}_{S_1}({\vc z})\right]
&=&{\bf E}\left[
         \tilde{Q}^{(1)}_{S_1[X_1^n,X_2^n,\cdots,X_{M_1}^n] }({\vc z})
        \right]
\nonumber\\
&=&{Q}_{S_1}({\vc z}),
\\
 {\bf E}\left[\tilde{Q}^{(2)}_{S_2}({\vc z})\right]
&=&{\bf E}\left[
         \tilde{Q}^{(2)}_{S_2[Y_1^n,Y_2^n,\cdots,Y_{M_2}^n]}({\vc z})
        \right]
\nonumber\\
&=&{Q}_{S_2}({\vc y}),
\\
 {\bf E}\left[\tilde{Q}^{(3)}_{S_3}({\vc z})\right]
&=&{\bf E}\left[
         \tilde{Q}_{
              3,S_3[X_1^nY_1^n,X_2^nY_2^n,\cdots,X_{M_1}^nY_{M_2}^n]
               }({\vc z})
        \right]
\nonumber\\
&=&{Q}_{S_3}({\vc z}).
\eeqa 

{\it Step 2 (Estimation of the Variational Distance):}
On the upper bound of $d(Q,\tilde{Q}_{i,S_i})$, we obtain the following
chain of inequalities:   
\beqa
\Wid& &d(Q,\tilde{Q}^{(i)})
\nonumber\\
\Wid&=&\sum_{{\lvc z}\in {\cal Z}^n}
         |\tilde{Q}^{(i)}({\vc z})-Q({\vc z})|
\nonumber\\
\Wid&=&\sum_{{\lvc z}\in {\cal Z}^n}
         |\tilde{Q}^{(i)}_{S_i}({\vc z})
          +\tilde{Q}^{(i)}_{S_i^c}({\vc z})-Q_{S_i}({\vc z})
          -Q_{S_i^c}({\vc z})|
\nonumber\\
\Wid&\leq&\sum_{{\lvc z}\in {\cal Z}^n}
         \left\{
           |\tilde{Q}^{(i)}_{S_i}({\vc z})-Q_{S_i}({\vc z})|
           +\tilde{Q}^{(i)}_{S_i^c}({\vc z})+{Q}_{S_i^c}({\vc z})
         \right\}
\nonumber\\
\Wid&=&\sum_{ {\lvc z}\in S_{i,Z}}
          |\tilde{Q}^{(i)}_{S_i}({\vc z})-Q_{S_i}({\vc z})|
         +\sum_{{\lvc z}\in S^c_{i,Z}}  \tilde{Q}^{(i)}_{S_i^c}({\vc z})
\label{eqn:ieq00}
\\
\Wid & &+\Av \left[{\vc 1}_{S^c}(X^n,Y^n,Z^n)\right]\,.
\nonumber
\eeqa
Next we evaluate the first and second terms in the right member of   
(\ref{eqn:ieq00}).
For $i=1,2,3$, set 
\beqno
\Lambda_i&\defeq &
\sum_{ {\lvc z}\in S^c_{i,Z} } \tilde{Q}^{(i)}_{S_i^c}({\vc z}),
\Phi_i\defeq \sum_{{\lvc z}\in S_{i,Z} }
|\tilde{Q}^{(i)}_{S_i}({\vc z})-Q_{S_i}({\vc z})|\,.
\eeqno
We first observe that 
\beq
{\bf E}[\Lambda_i]=\Av \left[{\vc 1}_{S_i^c}(X^n,Y^n,Z^n)\right],
\mbox{ }i=1,2,3.\label{eqn:eq000z}
\eeq
Next we derive upper bounds of $\Phi_i, i=1,2,3$.
We first derive an upper bound of $\Phi_1$. Observe that  
\beqno
\hspace*{-3mm}& &\Phi_1
\nonumber\\
\hspace*{-3mm}&\leq& \sum_{{\lvc y}\in S_{1,Y}}
\hspace*{-3mm}P_{Y^n}({\vc y})
\nonumber\\
\hspace*{-3mm}&  &   \times \hspace*{-2mm}
         \sum_{{\lvc z}\in S_{1,Z|Y}({\lvc y})}
         \hspace*{-2mm}
         |\tilde{P}^{(1)}_{Z^n|Y^n,S_1}({\vc z}|{\vc y}) 
               -{P}_{Z^n|Y^n,S_1}({\vc z}|{\vc y})| 
\,.
\eeqno
Applying the Cauchy-Schwartz inequality and 
using the concavity of $\sqrt{x}$, we have
\beqa
\hspace*{-4mm}& &\Phi_1
\nonumber\\
\hspace*{-2mm}&\leq& \sum_{{\lvc y}\in S_{1,Y}}P_{Y^n}({\vc y})
        \times \left\{P_{Z^n|Y^n}(S_{1,Z|Y}({\vc y})|{\vc y})\right\}^{1/2}  
\nonumber\\
\hspace*{-4mm}& &\times \left\{
\hspace*{-6mm}\sum_{\hspace*{6mm}{\lvc z}\in S_{1,Z|Y}({\lvc y})}
\hspace*{-8mm}\frac{
      \left\{
             \tilde{P}^{(1)}_{Z^n|Y^n,S_1
                                     }({\vc z}|{\vc y})
              -P_{Z^n|Y^n,S_1}({\vc z}|{\vc y})
      \right\}^2}
      {P_{Z^n|Y^n}({\vc z}|{\vc y})}    
      \right\}^{1/2}   
\nonumber\\
\hspace*{-4mm}&\leq&
\sum_{{\lvc y}\in S_{1,Y}}P_{Y^n}({\vc y})
\nonumber\\
\hspace*{-4mm}& &\times \left\{
\hspace*{-6mm}\sum_{\hspace*{6mm}{\lvc z}\in S_{1,Z|Y}({\lvc y})}
\hspace*{-8mm}\frac{
      \left\{
             \tilde{P}^{(1)}_{Z^n|Y^n,S_1
                                     }({\vc z}|{\vc y})
              -P_{Z^n|Y^n,S_1}({\vc z}|{\vc y})
      \right\}^2}
      {P_{Z^n|Y^n}({\vc z}|{\vc y})}    
      \right\}^{1/2}   
\nonumber\\
\hspace*{-4mm}&\leq& \left\{ P_{Y^n}(S_{1,Y}) \right\}^{1/2}
\nonumber\\
\hspace*{-4mm}& &\times \HUgel \sum_{{\lvc y}\in S_{1,Y}}P_{Y^n}({\vc y})
\nonumber\\
\hspace*{-4mm}& &\times
\hspace*{-6mm}\sum_{\hspace*{6mm}{\lvc z}\in S_{1,Z|Y}({\lvc y})}
\hspace*{-8mm}\frac{
      \left\{
             \tilde{P}^{(1)}_{Z^n|Y^n,S_1
                                     }({\vc z}|{\vc y})
              -P_{Z^n|Y^n,S_1}({\vc z}|{\vc y})
      \right\}^2}
      {P_{Z^n|Y^n}({\vc z}|{\vc y})}    
      \HUger^{1/2}   
\nonumber\\
\hspace*{-4mm}&\leq&
\HUgel\sum_{({\lvc z},{\lvc y})\in S_{1,ZY}}P_{Y^n}({\vc y})
\nonumber\\
& &\qquad\times
      \frac{
      \left\{
             \tilde{P}^{(1)}_{Z^n|Y^n,S_1
                                     }({\vc z}|{\vc y})
              -P_{Z^n|Y^n,S_1}({\vc z}|{\vc y})
      \right\}^2}{P_{Z^n|Y^n}({\vc z}|{\vc y})}    
      \HUger^{1/2}   
\label{eqn:ieq10} 
\eeqa
Taking expectation of both sides of (\ref{eqn:ieq10}) and 
using Jensen's inequality, we have  
\beqno
\hspace*{-4mm}& &{\bf E}[\Phi_1]
\nonumber\\
\hspace*{-4mm}&\leq&
      \left\{\hspace*{-5mm}
\sum_{\hspace*{5mm}({\lvc z},{\lvc y})\in S_{1,ZY}}
\hspace*{-5mm}P_{Y^n}({\vc y})      
      \frac{
      \mbox{\bf Var}\left[
      \tilde{P}^{(1)}_{Z|Y,S_1
                 }({\vc z}|{\vc y})
      \right]}{ P_{Z^n|Y^n}({\vc z}|{\vc y}) }
      \right\}^{1/2}.
\label{eqn:ieqphi-1}
\eeqno
In a manner quite similar to the above argument we obtain
\beqno
\hspace*{-4mm}& &{\bf E}[\Phi_2]
\nonumber\\
\hspace*{-4mm}&\leq&
      \left\{\hspace*{-5mm}
\sum_{\hspace*{5mm}({\lvc z},{\lvc x})\in S_{2,ZX}}
\hspace*{-5mm}P_{X^n}({\vc x})      
      \frac{
      \mbox{\bf Var}\left[
      \tilde{P}^{(2)}_{Z|X,S_2
                 }({\vc z}|{\vc x})
      \right]}{ P_{Z^n|X^n}({\vc z}|{\vc x}) }
      \right\}^{1/2}.
\label{eqn:ieqphi-2}
\eeqno
Next, we derive an upper bound of $\Phi_3$. 
Applying the Cauchy-Schwartz inequality, we have
\beqa
\Phi_3
&\leq&\left\{\sum_{{\lvc z}\in S_{3,Z}}
{Q}_{}({\vc z})\right\}^{1/2}  
\nonumber\\
&    &\times\left\{
      \sum_{{\lvc z}\in S_{3,Z}}
      \frac{
      \left\{
             \tilde{Q}^{(3)}_{ S_3
                              }({\vc z})
             -Q_{S_3}({\vc z})
      \right\}^2}{ Q({\vc z})}
      \right\}^{1/2}   
\nonumber\\
&\leq &\left\{
      \sum_{{\lvc z}\in S_{3,Z}}
      \frac{
      \left\{
             \tilde{Q}^{(3)}_{ S_3
                              }({\vc z})
             -Q_{S_3}({\vc z})
      \right\}^2}{ Q({\vc z})}
      \right\}^{1/2}   
\label{eqn:ieq10a} 
\eeqa
Taking expectation of both sides of (\ref{eqn:ieq10a}) and 
using Jensen's inequality, we have  
\beqa
{\bf E}[\Phi_3]
&\leq&\left\{
\sum_{{\lvc z}\in S_{3,Z}}
      \frac{
      \mbox{\bf Var}\left[
      \tilde{Q}^{(3)}_{ S_3
             }({\vc z})
      \right]}{Q({\vc z})}
      \right\}^{1/2}\,.
\label{eqn:ieqphi-3}
\eeqa
{\it Step 3(Computation of the Variances)  
:} 
Observe that
\beqa
& &\left\{\tilde{P}^{(1)}_{Z^n|Y^n,S_1}({\vc z}|{\vc y})\right\}^2
\nonumber\\
&=&\frac{1}{M_1^2}\sum_{j=1}^{M_1}
   \sum_{{\lvc x}\in S_{1,X|ZY}({\lvc z},{\lvc y})} 
      \left[W^n({\vc z}|{\vc x},{\vc y})\right]^2
       \chi_{{\lvc x}}(X_j^n)
\nonumber\\
& &
+\frac{1}{M_1^2}\sum_{j\ne j^{\prime}}
\sum_{{\lvc x}\in S_{1,X|ZY}({\lvc z},{\lvc y})}
\sum_{{\lvc x}^{\prime}\in S_{1,X|ZY}({\lvc z},{\lvc y})}
\nonumber\\
& &\qquad \times
      W^n({\vc z}|{\vc x},{\vc y})
      W^n({\vc z}|{\vc x}^{\prime},{\vc y})
      \chi_{{\lvc x}}(X_j^n)
      \chi_{{\lvc x}^{\prime}}(X_{j^{\prime}}^n)\,.
\label{eqn:ieq03}
\eeqa
Taking expectation of both sides of (\ref{eqn:ieq03}), we obtain
\beqno
& & {\bf E}\left[\left\{
           \tilde{P}^{(1)}_{Z^n|Y^n,S_1}({\vc z}|{\vc y})
           \right\}^2\right]
\nonumber\\
&\leq&\frac{1}{M_1}\sum_{{\lvc x}\in S_{1,X|ZY}({\lvc z},{\lvc y})} 
      \left[W^n({\vc z}|{\vc x},{\vc y})\right]^2 P_{X^n}({\vc x})
\nonumber\\
& & +\left\{P_{Z^n|Y^n,S_1}({\vc z}|{\vc y})\right\}^2.
\label{eqn:ieq04}
\eeqno
Thus, we have
\beqno
& &\mbox{\bf Var}\left[
                 \tilde{P}^{(1)}_{Z^n|Y^n,S_1}({\vc z}|{\vc y})
                 \right]
\nonumber\\
&\leq&\frac{1}{M_1}
   \sum_{{\lvc x}\in S_{1,X|ZY,\gamma}({\lvc z},{\lvc y})} 
      \left[W^n({\vc z}|{\vc x},{\vc y})\right]^2 P_{X^n}({\vc x})\,.
\label{eqn:ieq05}
\eeqno
From the above inequality and (\ref{eqn:ieqphi-1}), 
we obtain
\beqa
& &{\bf E}\left[\Phi_1\right]
\nonumber\\
&\leq &\hspace*{-1mm}\left\{\hspace*{-5.5mm}      
     \sum_{\hspace*{5mm}({\lvc z},{\lvc y})\in S_{1,ZY}}
     \hspace*{-8mm}      
     \sum_{\hspace*{8mm}{\lvc x}\in S_{1,X|ZY}({\lvc z},{\lvc y})} 
     \hspace*{-11mm}      
     P_{Y^n}({\vc y})
     \frac{
     \left[W^n({\vc z}|{\vc x},{\vc y})\right]^2\hspace*{-1mm}
      P_{X^n}({\vc x})
      }{M_1 P_{Z^n|Y^n}({\vc z}|{\vc y})}
      \hspace*{-1mm}\right\}^{1/2}
\nonumber\\
&= & \Hugel 
     \sum_{({\lvc x},{\lvc y}, {\lvc z})\in S_1} 
     \exp\left\{-n\left[
             R_1-{\ts \frac{1}{n}}i_{X^nY^nZ^n}({\vc x};{\vc z}|{\vc y})
             \right]^{\empty}
          \right\}
\nonumber\\
& &\qquad\qquad\qquad\qquad
   \times W^n({\vc z}|{\vc x},{\vc y})
    P_{X^n}({\vc x})P_{Y^n}({\vc y})
    \Huger^{1/2} 
\nonumber\\
&=&\sqrt{\zeta_{n,1,S_1}\Aga}\,.
\label{eqn:ieq08}
\eeqa
In a manner quite similar to the above argument we obtain   
\beq
{\bf E}\left[\Phi_2\right]
\leq\sqrt{\zeta_{n,2,S_2}\Agb}\,.
\label{eqn:ieq08b}
\eeq
Next, we compute ${\bf Var}[\tilde{Q}^{(3)}_{S_3}({\vc z})]$.
\renewcommand{\Wid}{\hspace*{-7mm}}
Observe that
\beqa
\Wid& &\left\{\tilde{Q}^{(3)}_{S_3} ({\vc z})\right\}^2
\nonumber\\
\Wid&=&\frac{1}{M_1^2M_2^2}\sum_{j=1}^{M_1}\sum_{k=1}^{M_2}
   \sum_{({\lvc x},{\lvc y})\in S_{3,XY|Z}({\lvc z})} 
\nonumber\\
\Wid&&\times
               \left[W^n({\vc y}|{\vc x},{\vc y})\right]^2
               \chi_{{\lvc x}}(X_j^n)\chi_{{\lvc y}}(Y_k^n)
\nonumber\\
\Wid&&+\frac{1}{M_1^2M_2^2}\sum_{j\ne j^{\prime}}\sum_{k=1}^{M_2}
     \sum_{\scs ({\lvc x},{\lvc y})\in S_{3,XY|Z}({\lvc z})
               \atop{\scs ({\lvc x}^{\prime},{\lvc y})\in S_{3,XY|Z}({\lvc z})
                }
           } 
\nonumber\\
\Wid&&\times   W^n({\vc z}|{\vc x},{\vc y})
               W^n({\vc z}|{\vc x}^{\prime},{\vc y})
               \chi_{{\lvc x}}(X_j^n)
               \chi_{{\lvc x}^{\prime}}(X_{j^{\prime}}^n)
               \chi_{{\lvc y}}(Y_k^n)
\nonumber\\
\Wid&&+\frac{1}{M_1^2M_2^2}\sum_{j=1}^{M_1}\sum_{k\ne k^{\prime}}
     \sum_{\scs ({\lvc x},{\lvc y})\in S_{3,XY|Z}({\lvc z})
               \atop{\scs ({\lvc x},{\lvc y}^{\prime})
                           \in S_{3,XY|Z}({\lvc z})
                }
           } 
\nonumber\\
\Wid&&\times  W^n({\vc z}|{\vc x},{\vc y})
              W^n({\vc z}|{\vc x},{\vc y}^{\prime})
                \chi_{{\lvc x}}(X_j^n)
                \chi_{{\lvc y}}(Y_k^n)
                \chi_{{\lvc y}^{\prime}}(Y_{k^{\prime}}^n)
\nonumber\\
\Wid&&+\frac{1}{M_1^2M_2^2}\sum_{j\ne j^{\prime}}\sum_{k\ne k^{\prime}}
     \sum_{\scs ({\lvc x},{\lvc y})\in S_{3,XY|Z}({\lvc z})
               \atop{\scs ({\lvc x}^{\prime},{\lvc y}^{\prime})
                          \in S_{3,XY|Z}({\lvc z})
                }
           } 
\nonumber\\
\Wid&&   \times W^n({\vc z}|{\vc x},{\vc y})
                W^n({\vc z}|{\vc x}^{\prime},{\vc y}^{\prime})
\nonumber\\
\Wid&&\times    \chi_{{\lvc x}}(X_j^n)
                \chi_{{\lvc x}^{\prime}}(X_{j^{\prime}}^n)
                \chi_{{\lvc y}}(Y_k^n)
                \chi_{{\lvc y}^{\prime}}(Y_{k^{\prime}}^n).
\label{eqn:ieq03a}
\eeqa
Taking expectation of both sides of (\ref{eqn:ieq03a}), we obtain
\renewcommand{\Wid}{\hspace*{-0mm}}
\beqno
\Wid& & {\bf E}\left[\left\{
           \tilde{Q}^{(3)}_{S_3}({\vc z})
           \right\}^2\right]
\nonumber\\
\Wid&\leq&\frac{1}{M_1M_2}
      \sum_{({\lvc x},{\lvc y})
              \in S_{3, XY|Z}({\lvc z})
            }
\left[W^n({\vc z}|{\vc x},{\vc y})\right]^2 
             P_{X^n}({\vc x})P_{Y^n}({\vc y})
\nonumber\\
\Wid&&+\frac{1}{M_2}
      \sum_{{\lvc y}\in S_{3,Y|Z}({\lvc z})} 
      \sum_{\scs {\lvc x},{\lvc x}^{\prime} 
             \in S_{3, X|YZ}({\lvc z},{\lvc y})
           }
\nonumber\\
\Wid&&\times W^n({\vc z}|{\vc x},{\vc y})
             W^n({\vc z}|{\vc x}^{\prime},{\vc y})
             P_{X^n}({\vc x})
             P_{X^n}({\vc x}^{\prime})
             P_{Y^n}({\vc y})
\nonumber\\
\Wid&&+\frac{1}{M_1}
      \sum_{{\lvc x}\in S_{3,X|Z}({\lvc z})} 
      \sum_{\scs {\lvc y},{\lvc y}^{\prime} 
             \in S_{3, Y|ZX}({\lvc z},{\lvc x})
           }
\nonumber\\
\Wid&&\times W^n({\vc z}|{\vc x},{\vc y})
             W^n({\vc z}|{\vc x},{\vc y}^{\prime})
             P_{X^n}({\vc x})
             P_{Y^n}({\vc y})
             P_{Y^n}({\vc y}^{\prime})
\nonumber\\
\Wid&& +\left\{{Q}_{S_3}({\vc z})\right\}^2.
\label{eqn:ieq04a}
\eeqno
Thus, we have
\beqno
\Wid& &\mbox{\bf Var}\left[\tilde{Q}^{(3)}_{S_3}({\vc z})\right]
\nonumber\\
\Wid&\leq &\frac{1}{M_1M_2}
   \sum_{({\lvc x},{\lvc y})\in S_{3, XY|Z}({\lvc z})} 
      \left[W^n({\vc z}|{\vc x},{\vc y})\right]^2 
       P_{X^n}({\vc x})P_{Y^n}({\vc y})
\nonumber\\
\Wid&&+\frac{1}{M_2}
      \sum_{{\lvc y}\in S_{3, Y|Z}({\lvc z})} 
      \left[W^n({\vc z}|{\vc y})\right]^2 
       P_{Y^n}({\vc y})
\nonumber\\
\Wid&&+\frac{1}{M_1}
   \sum_{{\lvc x}\in S_{3, X|Z}({\lvc z})} 
      \left[W^n({\vc z}|{\vc x})\right]^2 
       P_{X^n}({\vc x})\,.
\label{eqn:ieq05a}
\eeqno
From the above inequality and (\ref{eqn:ieqphi-3}), 
we obtain
\beqa
& &{\bf E}\left[\Phi_3\right]
\nonumber\\
&\leq &
    \Hugel 
     \sum_{{\lvc z}\in S_{3,Z}}
     \sum_{({\lvc x},{\lvc y})\in S_{3,XY|Z}({\lvc z})} 
\hspace*{-2mm}     
      \frac{
     \left[W^n({\vc z}|{\vc x},{\vc y})\right]^2 
           P_{X^n}({\vc x})P_{Y^n}({\vc y})
      }{M_1M_2 Q({\vc z})}
\nonumber\\
& &+ \sum_{{\lvc z}\in S_{3,Z}}
     \sum_{{\lvc y}\in S_{3,Y|Z}({\lvc z})}
\hspace*{-2mm}      
     \frac{
     \left[W^n({\vc z}|{\vc y})\right]^2 P_{Y^n}({\vc y})
          }{M_2 Q({\vc z})}
\nonumber\\
&&+ \sum_{{\lvc z}\in S_{3,Z}}
     \sum_{{\lvc x}\in S_{3,X|Z}({\lvc z})}
\hspace*{-2mm}      
     \frac{
     \left[W^n({\vc z}|{\vc x})\right]^2 P_{X^n}({\vc x})
           }{M_1 Q({\vc z})}
    \Huger^{1/2} 
\nonumber\\
&=&\Hugel 
     \sum_{({\lvc x},{\lvc y},{\lvc z})\in S_{3}}
\hspace*{-2mm}      
     \exp\left\{-n\left[
             R_1+R_2 - {\ts \frac{1}{n}}i_{X^nY^nZ^n}({\vc x}{\vc y};{\vc z})
             \right]^{\empty}
          \right\}
\nonumber\\
& &\qquad\qquad\qquad\qquad
   \times W^n({\vc z}|{\vc x}{\vc y})
   P_{X^n}({\vc x})P_{Y^n}({\vc y})
\nonumber\\
&&\quad+ \sum_{({\lvc y},{\lvc z})\in S_{2,YZ}} 
        \exp\left\{-n\left[
             R_2- {\ts \frac{1}{n}}i_{Y^nZ^n}({\vc y};{\vc z})
             \right]^{\empty}
          \right\}
\nonumber\\
& &\qquad\qquad\qquad\qquad
   \times W^n({\vc z}|{\vc y})P_{Y^n}({\vc y})
\nonumber\\
&&\quad+ \sum_{({\lvc x},{\lvc z})\in S_{1,XZ}} 
     \exp\left\{-n\left[
             R_1-{\ts \frac{1}{n}}i_{X^nZ^n}({\vc x};{\vc z})
             \right]^{\empty}
          \right\}
\nonumber\\
& &\qquad\qquad\qquad\qquad
   \times W^n({\vc z}|{\vc x})P_{X^n}({\vc x})
   \Huger^{1/2} 
\nonumber\\
&=&\sqrt{\zeta_{n,3,S_3}\Ag}\,.
\label{eqn:ieq08a}
\eeqa
Set 
\beqno
\Theta_i &\defeq& \Av \left[{\vc 1}_{S_i^c}(X^n,Y^n,Z^n)\right]
+\sqrt{\zeta_{n,i,S_i}},\:i=1,2,3. 
\eeqno
From (\ref{eqn:eq000z}), (\ref{eqn:ieq08}), (\ref{eqn:ieq08b}), 
and (\ref{eqn:ieq08a}), we obtain
\beqno
&&
{\bf E}\left[\sum_{i=1,2,3}\Theta_i^{-1}(\Lambda_i+\Phi_i)\right]
\\
&=&\sum_{i=1,2,3}\Theta_i^{-1}
\left\{ {\bf E} [\Lambda_i]+{\bf E}[\Phi_i] \right\}
\\
&\leq &
\sum_{i=1,2,3}
\Theta_i^{-1}\left\{
\Av \left[{\vc 1}_{S_i^c}(X^n,Y^n,Z^n)\right]+\sqrt{\zeta_{n,i,S_i}}\right\}=3.
\eeqno
Then, there exists at least one deterministic maps 
$\resenc_i, i=1,2$ such that
$$
\sum_{i=1,2,3}\Theta_i^{-1}(\Lambda_i+\Phi_i)\leq 3,
$$
from which we have 
\beq
\Lambda_i+\Phi_i\leq 3\Theta_i, i=1,2,3.
\label{eqn:apa}
\eeq
From (\ref{eqn:ieq00}) and (\ref{eqn:apa}), we obtain
\beqno
&&d(Q,\tilde{Q}^{(i)})
\nonumber\\
&\leq& 4\Av\left[{\vc 1}_{S_i^c}(X^n,Y^n,Z^n)\right]
+3\sqrt{\zeta_{n,i,S_i}},\mbox{ }i=1,2,3,
\eeqno
completing the proof of Lemma \ref{lm:mlm}\,. 
\hfill\IEEEQED


\begin{thebibliography}{9}

\bibitem{ad}R. Ahlswede and G. Dueck,
``Identification via channels'' 
{\it IEEE Trans. Inform. Theory}, vol. 35, pp. 15-29, Jan. 1989. 

\bibitem{ad2}\underline{$\qquad$},
``Identification in the presence of feedback 
--A discovery of new capacity formulas'' 
{\it IEEE Trans. Inform. Theory}, vol. 35, pp. 30-36, Jan. 1989. 

\bibitem{hv1}T. S. Han and S. Verd\'u, 
``New results in the theory of identification via channels,''
{\it IEEE Trans. Inform. Theory}, vol. 38, pp. 14-25, Jan. 1992.  

\bibitem{hv2}\underline{$\qquad$},
``Approximation theory of output statistics,''
{\it IEEE Trans. Inform. Theory}, vol. 39, pp. 752-722, May 1993.  

\bibitem{BV}M. V. Burnashev ad S. Verd\'u, ``Measures separated in L1
metrics and ID codes,''
{\it Probl. Inform. Transm.}, vol. 30, no. 3, pp. 191-201, 1994. 

\bibitem{AZ} R. Ahlswede and Z. Zhang, 
``New directions in the theory of identification via channels,''
{\it IEEE Trans. Inform. Theory} vol. 41, pp. 1040-1050, July 1995. 

\bibitem{BB}L. A. Bassalygo and M.V. Burnashev, 
``Authentication, identification and pairwise separated measures,''
{\it Probl. Inform. Transm.}, vol. 32, no. 1, pp. 33-39, 1996. 

\bibitem{stb}Y. Steinberg, 
``New converses in the theory of identification via channels,''
{\it IEEE Trans. Inform. Theory}, vol. 44, pp. 984-998, May 1998.  

\bibitem{han}T. S. Han,
{\it Information-Spectrum Methods in Information Theory. (in Japanese)}  
Baifukan, Tokyo, 1998.

\bibitem{b2}M. V. Burnashev,
``On identification capacity of infinite alphabets or 
continuous-time channels,''
{\it IEEE Trans. Inform. Theory} vol. 46, pp. 2407-2414, Nov. 2000. 

\bibitem{stmr}Y. Steinberg and N. Merhav, 
``Identification in the presence of side information with application 
to watermarking,''
{\it IEEE Trans. Inform. Theory}, vol. 47, pp. 1410-1422, May 2001.  

\bibitem{hy}M. Hayashi,``General non-asymptotic and asymptotic 
formulas in channel resolvability and identification capacity 
and its application to wire-tap channel,'' 
{\it IEEE Trans. Inform. Theory}, vol. 52, no. 4, pp. 1562-1575, April 2006.

\bibitem{ohIdch}Y. Oohama, 
``Converse coding theorems for identification via channels,''    
{\it IEEE Trans. Inform. Theory}, vol. 59, pp. 744-759, Feb. 2013. 

\bibitem{VM}B. Verboven and E. C. van der Meulen, 
``Capacity bounds for identification via broad cast channels 
that are optimal for the deterministic broadcast channel,''    
{\it IEEE Trans. Inform. Theory}, vol. 36, pp. 1197-1205, Nov. 1990. 

\bibitem{AV} R. Ahlswede and B. Verboven, ``On identification via multiway
channels with feedback,'' 
{\it IEEE Trans. Inform. Theory} vol. 37, pp. 1519-1526, Nov. 1991. 

\bibitem{ohMacId}Y. Oohama, 
``Converse coding theorem for the identification 
via multiple access channels,''    
{\it IEEE Inform. Theory Workshop}, Bangalore, India, Oct. 20-25, 
pp. 155-158, 2002. 

\bibitem{stb2}Y. Steinberg, 
``Resolvability theory for the multiple-accsess channel,''
{\it IEEE Trans. Inform. Theory}, vol. 44, pp. 472-487, March 1998.  

\bibitem{hanMac} T. S. Han, ``An information spectrum approach 
to capacity theorems for the general multiple-access channel,'' 
{\it IEEE Trans. Inform. Theory} vol. 44, pp. 2773-2795, Nov. 1998. 

\bibitem{Ved}S. Verd\'u, 
``Multiple-access channels with memory with and without frame synchronism,'' 
{\it IEEE Trans. Inform. Theory}, vol 35, pp. 605-619, May 1989. 
\end{thebibliography}
\end{document}